\documentclass[sn-mathphys-num]{sn-jnl}

\usepackage{graphicx}%
\usepackage{multirow}%
\usepackage{amsmath,amssymb,amsfonts}%
\usepackage{amsthm}%
\usepackage[title]{appendix}%
\usepackage{xcolor}%
\usepackage{textcomp}%
\usepackage{manyfoot}%
\usepackage{booktabs}%
\usepackage{algorithm}%
\usepackage{algorithmicx}%
\usepackage{algpseudocode}%
\usepackage{listings}%
\usepackage{subfigure}
\DeclareUnicodeCharacter{02BC}{'}
\usepackage{tabularx}
\usepackage{float}
\usepackage{lmodern}

\begin{document}

\title[Article Title]{Fractal structure of Christ-Lee model}

\author*[1]{\fnm{Bo-Ming} \sur{Zhang}}\email{zhangboming2003@126.com}

\author[2]{\fnm{Yunguo} \sur{Jiang}}\email{jiangyg@sdu.edu.cn}
\equalcont{Corresponding author: jiangyg@sdu.edu.cn}

\affil*[1]{\orgdiv{SDU-ANU Joint Science College}, \orgname{Shandong University}, \orgaddress{\street{180 West Wenhua Road}, \city{Weihai}, \postcode{264209}, \state{Shandong Province}, \country{Peopleʼs Republic of China}}}

\affil[2]{\orgdiv{Shandong Provincial Key Laboratory of Optical Astronomy and Solar-Terrestrial Environment}, \orgname{Institute of Space Sciences, Shandong University}, \orgaddress{\street{180 West Wenhua Road}, \city{Weihai}, \postcode{264209}, \state{Shandong Province}, \country{Peopleʼs Republic of China}}}








\abstract{This paper numerically investigates the dynamical properties of kink and antikink collisions in the Christ-Lee model in the regime of $\epsilon$  approaching the $\phi^4$ theory. With given  $\epsilon$ and the initial velocity $V_{in}$, we exhibiting the formation of bion, scattering and $n$-bounce states.  Additionally, we show the self-similar fractal structures in the plot of $V_{out}-V_{in}$ with given $\epsilon$. Specially, we find the fractal structure in the plot of $ V_{out}$ versus $\epsilon$, which is not reported previously. We computes the Box-counting dimension for these fractal structures. We find that the Box-dimension is positively correlated with $\epsilon$, and approaches to Hausdorff dimension of the Sierpinski triangle when $\epsilon$ is sufficiently large.}


\maketitle

\section{Introduction}\label{sec1}

Recently, polynomial potential models have garnered considerable attention, including the $\phi^6$~\cite{dorey2011kink, gani2014kink, weigel2014kink} and $\phi^8$~\cite{bazeia2019kink, belendryasova2019scattering, gani2015kink} theories. However, the Christ-Lee (CL) model~\cite{christ1975quantum}, which exhibits unique properties by bridging the gap between the $\phi^4$ and $\phi^6$ theories, converging to the $\phi^6$ theory as $\epsilon$ approaches infinitesimal and to the $\phi^4$ theory as $\epsilon$ approaches infinity, has been rarely studied since its inception. The one study in the literature~\cite{dorey2023collisions} describes that  the CL model's \(\phi^6\)-like static solution reveals an inner structure as a pair of weakly bound subkinks when \(\epsilon\) approaches zero, and there exists a false vacuum at \(\phi = 0\), leading to the existence of unstable lumps, which complicates the soliton collision process.

An essential aspect of understanding soliton interactions is the exploration of fractal structures. Fractal structures, characterized by self-similarity across various scales, represent a distinctive form of geometric arrangement. One of key features is the fractal dimension, which quantifies the complexity and geometric properties of these structures. In recent years, there has been growing interest in studying fractal structures associated with different theories.  Levkov, Maslov, and Nugaev \cite{levkov2020chaotic} demonstrated that, in a one-dimensional sine-Gordon model with an external Dirac comb potential, the field values of stable solitons form a fractal structure, and the Box-counting dimension of this fractal was showed. Additionally, the distribution of field values within these fractals is closely related to the metric entropy of the corresponding mechanical system. The work by Anninos, Oliveira, and Matzner in~\cite{anninos1991fractal} revealed that, in the scalar $\lambda$($\varphi^2$-1)$^2$ theory, the fractal structure arises from kink-antikink collisions, and the self-similarity of this structure was elucidated. Moreover, this study calculated the box-counting dimension of the fractal structure and demonstrated the chaotic nature of the bounce state by computing the maximum Lyapunov exponent. Similarly, the fractal structure of the Christ-Lee model is also an interesting topic that requires in-depth investigation.

In this paper, we consider the behavior of the  CL model  as $\epsilon$ goes beyond $1/\sqrt{2}$ and to infinity, which is close to the $\phi^4$ theory. We use a fourth-order Runge-Kutta method in section~\ref{sec:three} to solve the PDE of soliton collisions for CL models under different $\epsilon$, and show the rich collision phenomena. In section~\ref{sec:four}, we present the fractal structure in  the $V_{\text{out}}-V_{\text{in}}$ plot for the CL model, along with the self-similarity of its unique $V_{\text{out}}-\epsilon$ fractal structure. We also calculate the Box-counting dimension and explore the impact of $\epsilon$ on the fractal structure and its Hausdorff dimension. Our study is meaningful to understand the formation of the fractal structure in a unique view point.

\section{Christ-Lee model}
We consider a model with the Lagrangian density that discussed in ~\cite{christ1975quantum}

    \begin{equation}
    \label{eq:1}
        \mathcal{L} = \frac{1}{2}\phi_t^2 - \frac{1}{2}\phi_x^2 - U(\phi ,\epsilon),
    \end{equation} 
where the subscript $t$ and $x$ denote the derivation with time and space respectively. The scalar potential is 
    \begin{equation}
        U(\phi ,\epsilon) = \frac{1}{2(1+\epsilon^2)}(\epsilon^2+\phi^2)(1-\phi^2)^2.
    \end{equation}

    \begin{figure}[htbp]
    \centering
    \subfigure[]{
        \includegraphics[width=.4\textwidth]{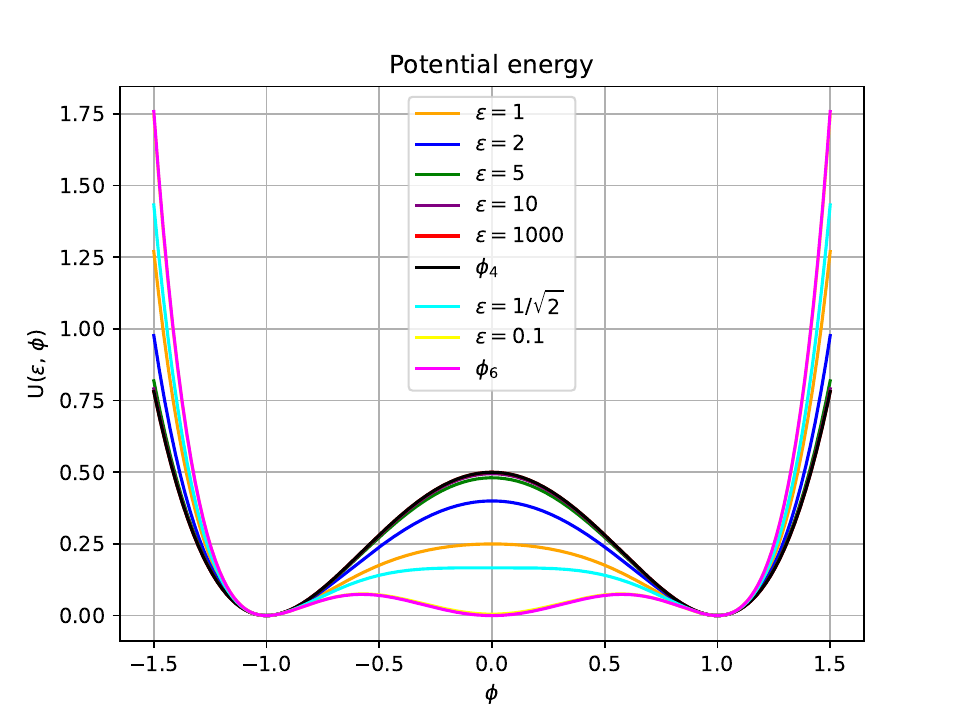}
        \label{fig:a-a}
    }
    \qquad
    \subfigure[]{
        \includegraphics[width=.4\textwidth]{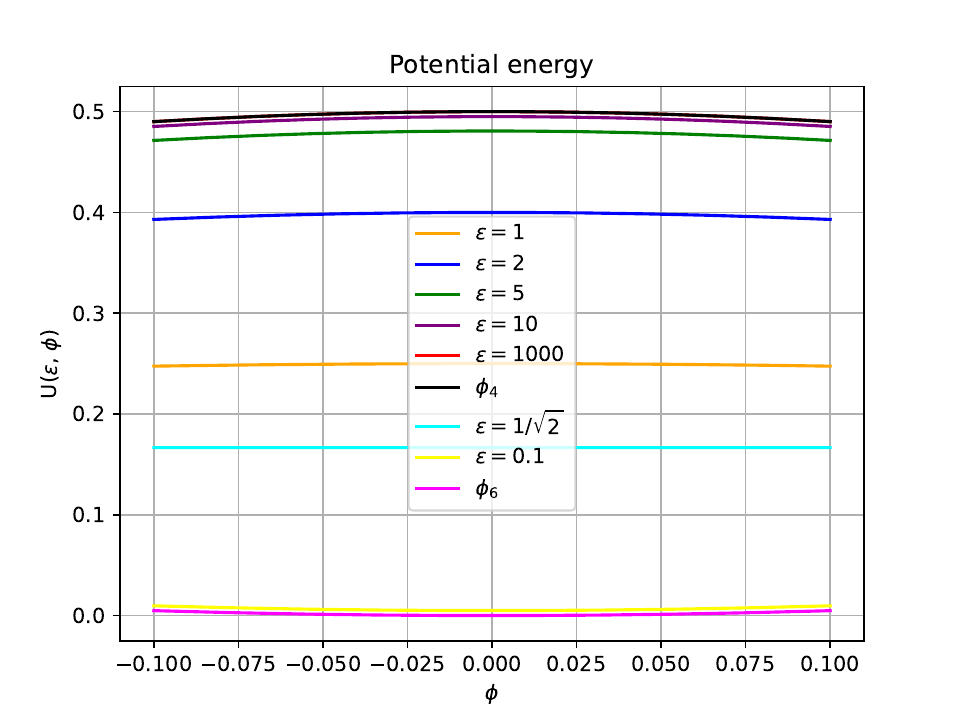}
        \label{fig:a-b}
    }
    \caption{\centering The global and local potential energy. }
    \label{fig:a}
    \end{figure}
    
The potential energy corresponding to different $\epsilon$ is shown in Figure~\ref{fig:a}. It has been found that $\epsilon = 1/\sqrt{2}$ is the critical point at which the convexity of the potential energy at the inflection point $\phi=0$ changes. From Figure \ref{fig:a-a}, it is evident that as $\epsilon$ approaches zero, the potential field converges to the triply-degenerate $\phi^6$ model. Conversely, as $\epsilon$ approaches infinity, the potential field converges to the well-known $\phi^4$ model. For $\epsilon \in (\frac{1}{\sqrt{2}}, +\infty)$, the potential has two local minima at $\phi = -1$ and $1$.  Additionally, The Figure \ref{fig:a-b} indicates the potential at the range $\phi \in (-0.1, 0.1)$ and reveals that when $\epsilon$ is greater than or equal to 5, the potential energy is already very close to the $\phi^4$ model.

\subsection{Static kink solutions}

\begin{figure}[htbp]
    \centering
    \includegraphics[width=0.7\textwidth]{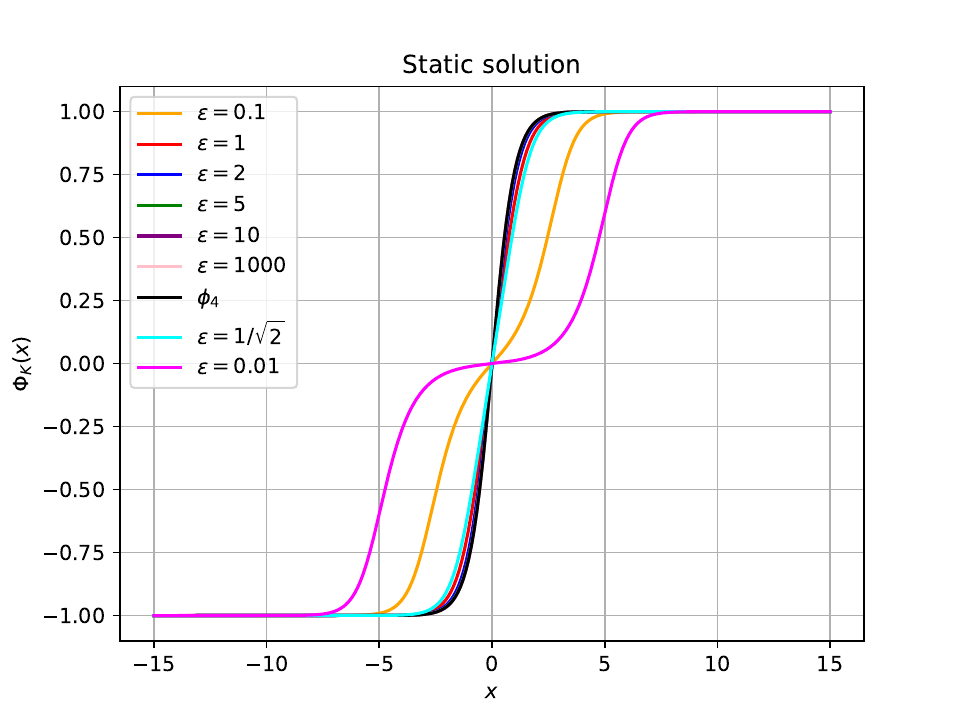}
    \caption{\centering The kink solution.}
    \label{fig:b}
\end{figure}

For  $\epsilon > 0$, one kink solution interpolating between the vacua at -1 and 1 is presented analytically 
    \begin{equation} \label{eq:phik}
        \Phi _K(x) = \frac{\epsilon \sinh(x)}{\sqrt{1+\epsilon^2 \cosh^2(x)}}.
    \end{equation}
In the Figure~\ref{fig:b}, we  show the static kink solutions for different values of $\epsilon$.
When the value of $\epsilon$ is greater than or equal to 5, the kink solution closely approximates the $\phi^4$ solution, as shown in Figure~\ref{fig:b}.

\section{Numerical set-up and results}\label{sec:three}

\subsection{Set up}

In this work, we use numerical methods to solve the dynamic equations of kink-antikink-pair collisions. 
From the Lagriangian in Equation~\ref{eq:1},  the Euler equation of the system is given as    
    \begin{equation}
    \label{eq:2}
        \frac{\partial^2 \phi}{\partial t^2} - \frac{\partial^2 \phi}{\partial x^2} + \frac{dU}{d\phi} = 0.
    \end{equation}   
This partial differential equation (PDE) is solved on a discrete spatial grid with periodic boundary conditions, where the total length is $2L = 50$.  The grid bin is $\Delta x =0.1$ in the simulation. 
Additionally, the time interval is chosen to be $\Delta$t = 0.05, and the total time step is $T_{tot} = 500$. We use a fourth-order Runge-Kutta method to perform the code to solve PDE.

The initial data in our simulation represents a kink and an antikink structure moving towards each other with the same incoming velocity $V_{\text{in}}$  at $t = 0$ with initial positions at 10 and -10, respectively. This is accomplished through the configuration formula
    \begin{equation}
         \Phi (x) = \Phi_{K}(x+x_0) + \Phi_{\bar{K}}(x-x_0) -1,
    \end{equation}
where $\Phi_{K}$ takes the form of Equation \ref{eq:phik}, and $\Phi_{\bar{K}}$ denotes the antikink which satisfies $\Phi_{\bar{K}}(x)=\Phi_{K}(-x)$. 
To numerically measure the final velocity and the escaping time, we define a truncated time $T_{\rm tru}$.
Normally, $T_{\rm tru}$ is the time when  the kink and antikink  reach the spatial boundary. When $T_{\rm tru}$ is less than the total time $ T_{\rm tot}$, we redefine the time interval to be $(0, T_{\rm tru})$.

\subsection{Kinetic results}

We use 3D images to illustrate the collision process for the Christ-Lee model. Two initial velocities $V_{\text{in}}=0.33$ and $0.06$ are chosen to manifest the scatter and bion states, respectively. We also vary $\epsilon$ to show how this parameter affects the collision dynamics. The values of  $\epsilon$ is chosen to be greater than $1/\sqrt{2}$ to present cases close to $\phi^4$ theory.

\begin{figure}[htbp]
\centering

\begin{tabular}{cccc} 
\includegraphics[width=.224\textwidth]{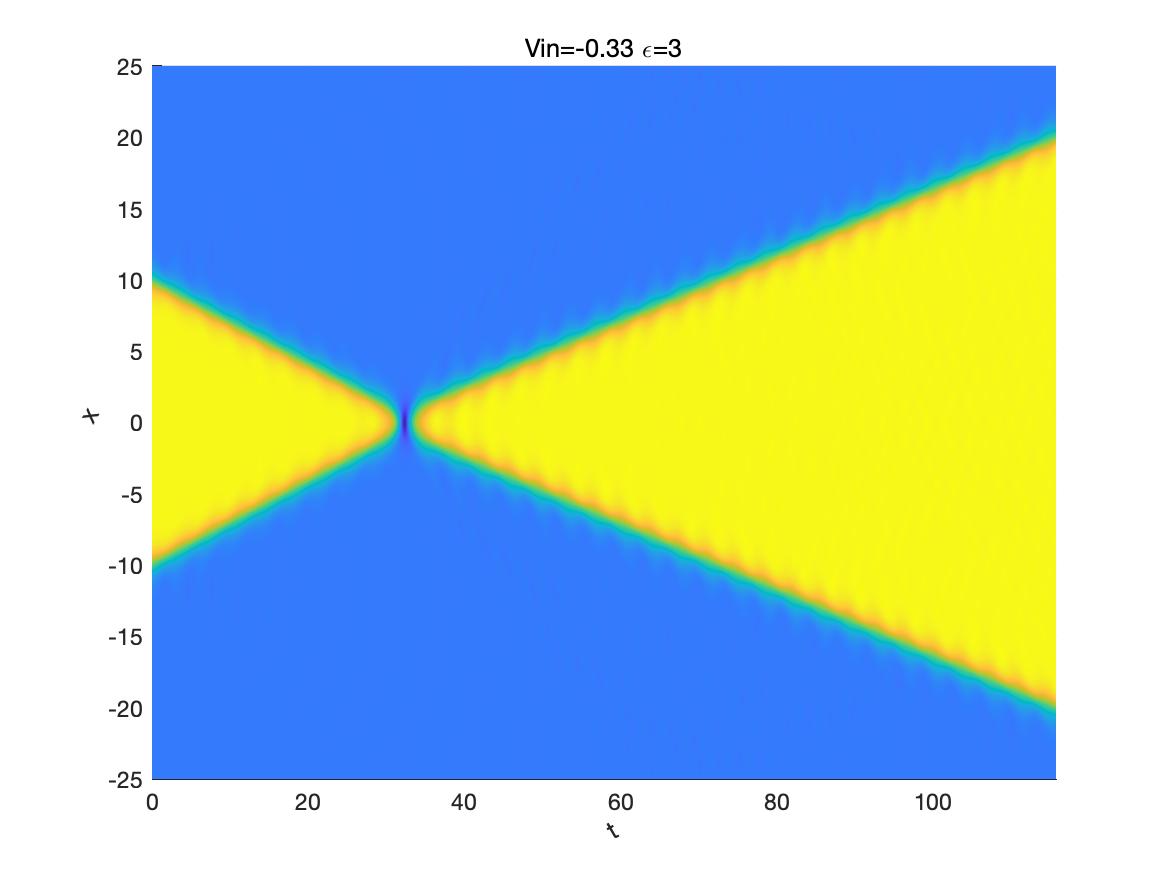} &
\hspace{-1em}
\includegraphics[width=.224\textwidth]{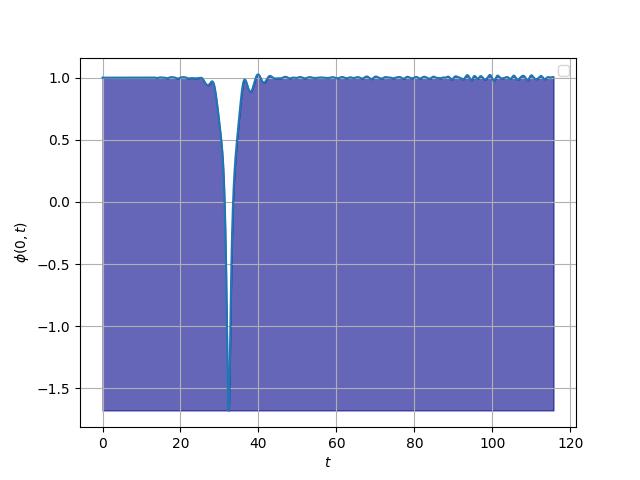} &
\hspace{-1em}
\includegraphics[width=.224\textwidth]{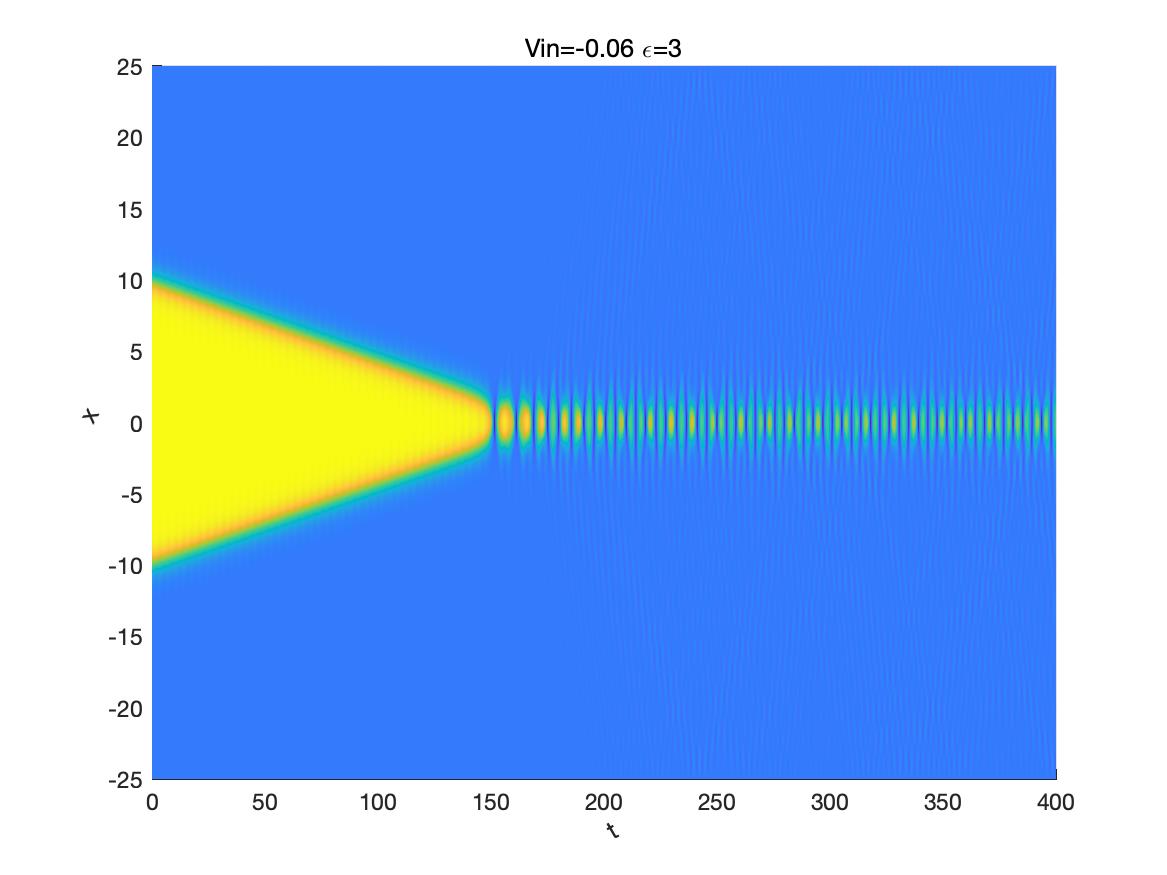} &
\hspace{-1em}
\includegraphics[width=.224\textwidth]{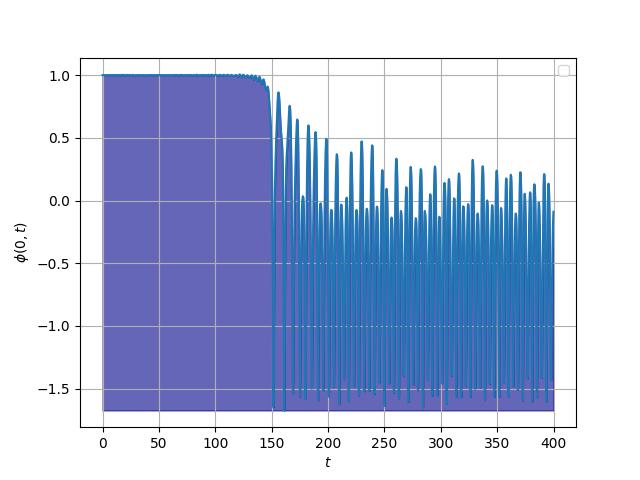} \\
\includegraphics[width=.224\textwidth]{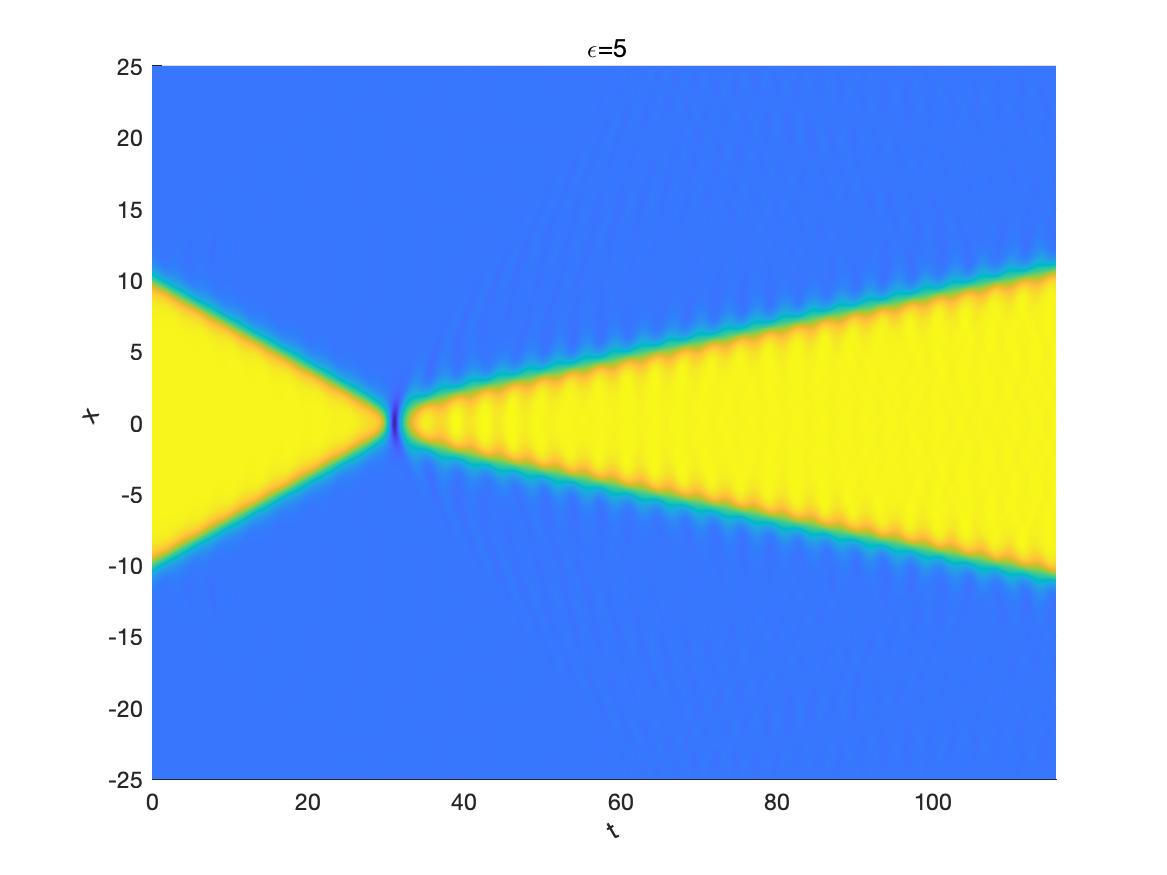} &
\hspace{-1em}
\includegraphics[width=.224\textwidth]{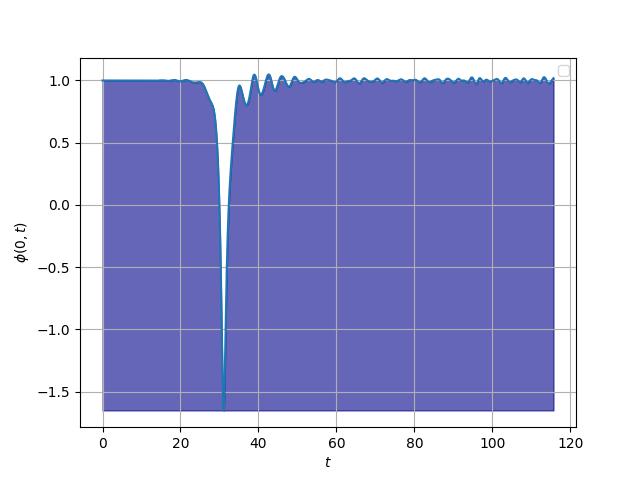} &
\hspace{-1em}
\includegraphics[width=.224\textwidth]{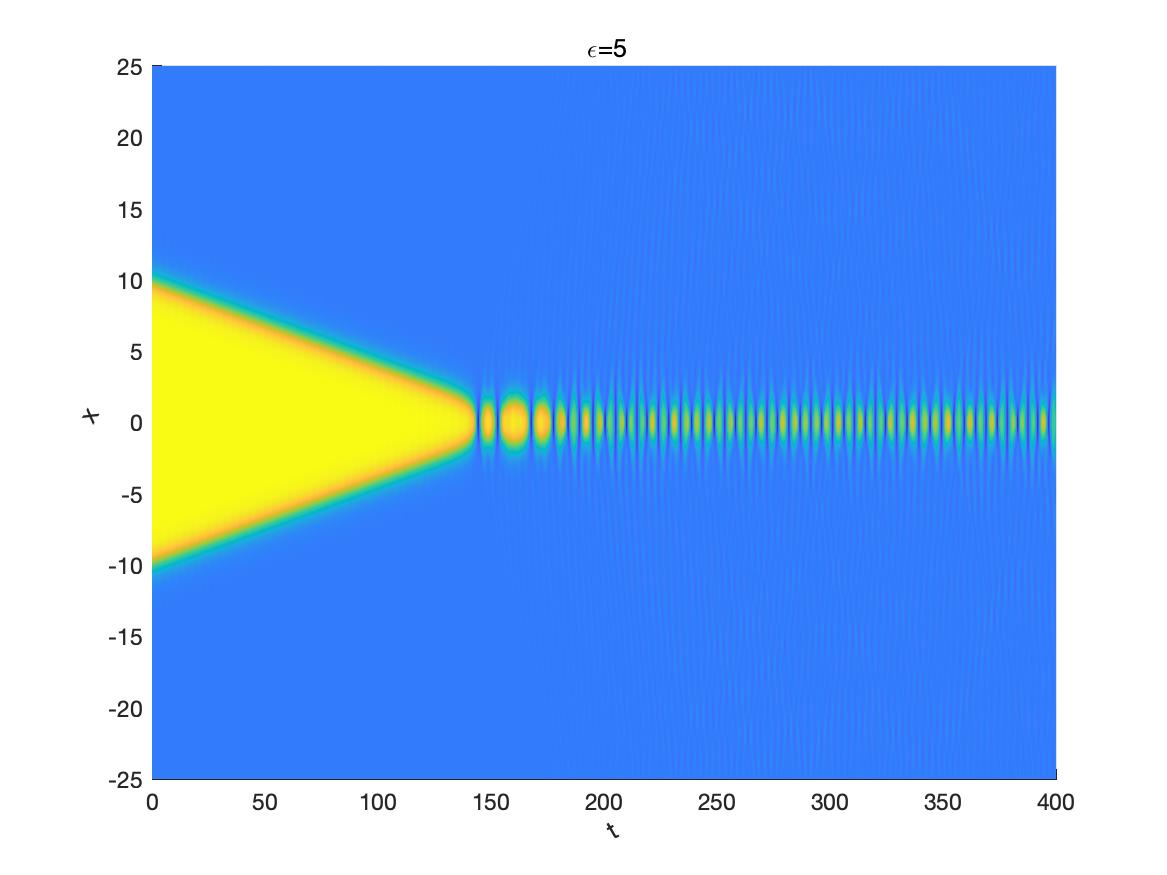} &
\hspace{-1em}
\includegraphics[width=.224\textwidth]{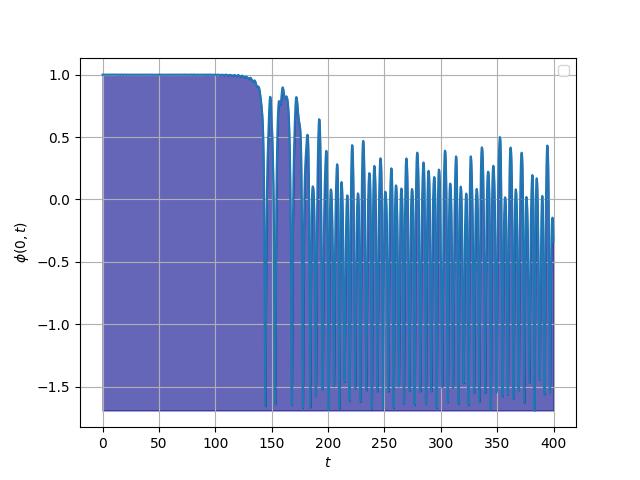} \\
\includegraphics[width=.224\textwidth]{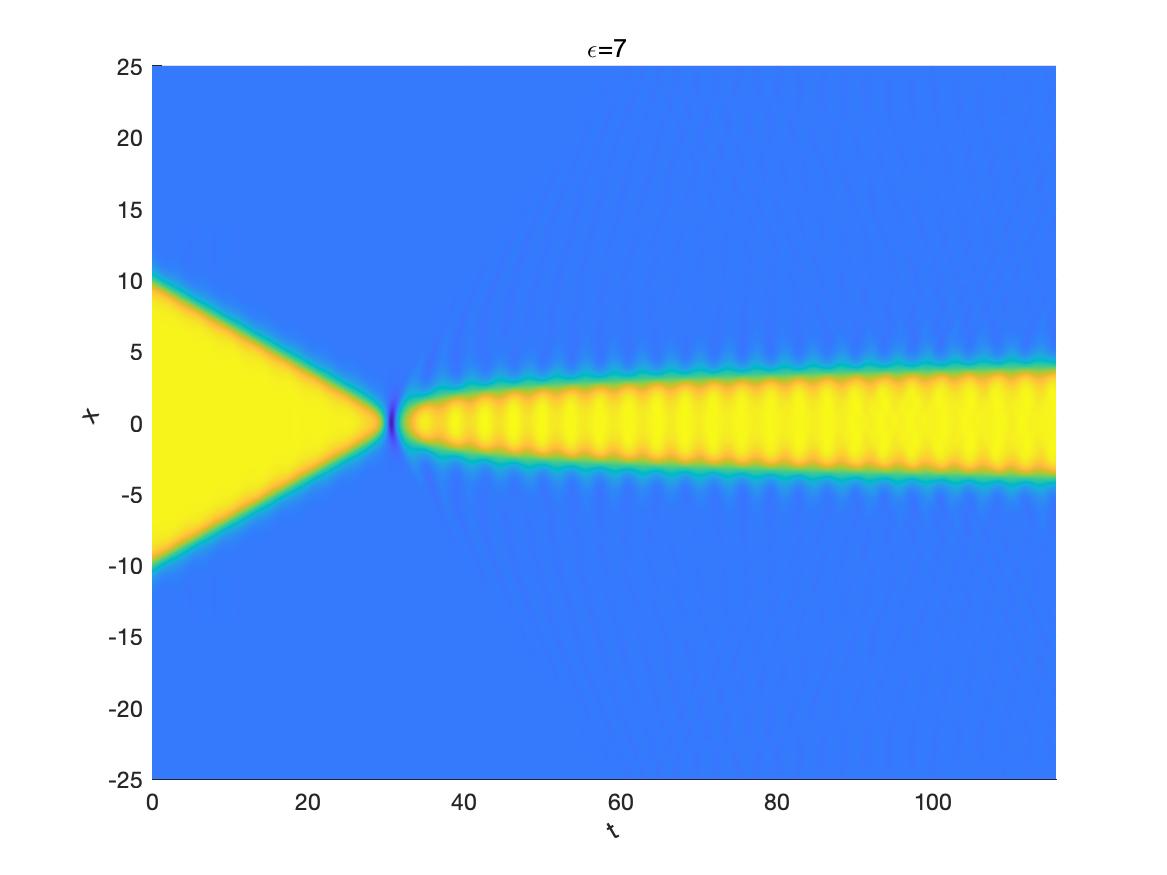} &
\hspace{-1em}
\includegraphics[width=.224\textwidth]{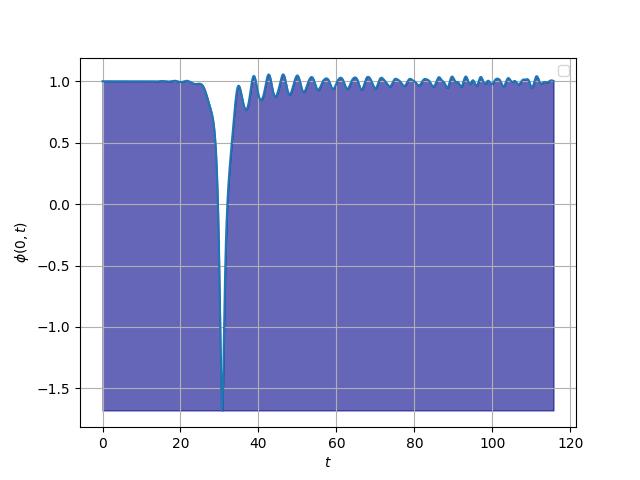} &
\hspace{-1em}
\includegraphics[width=.224\textwidth]{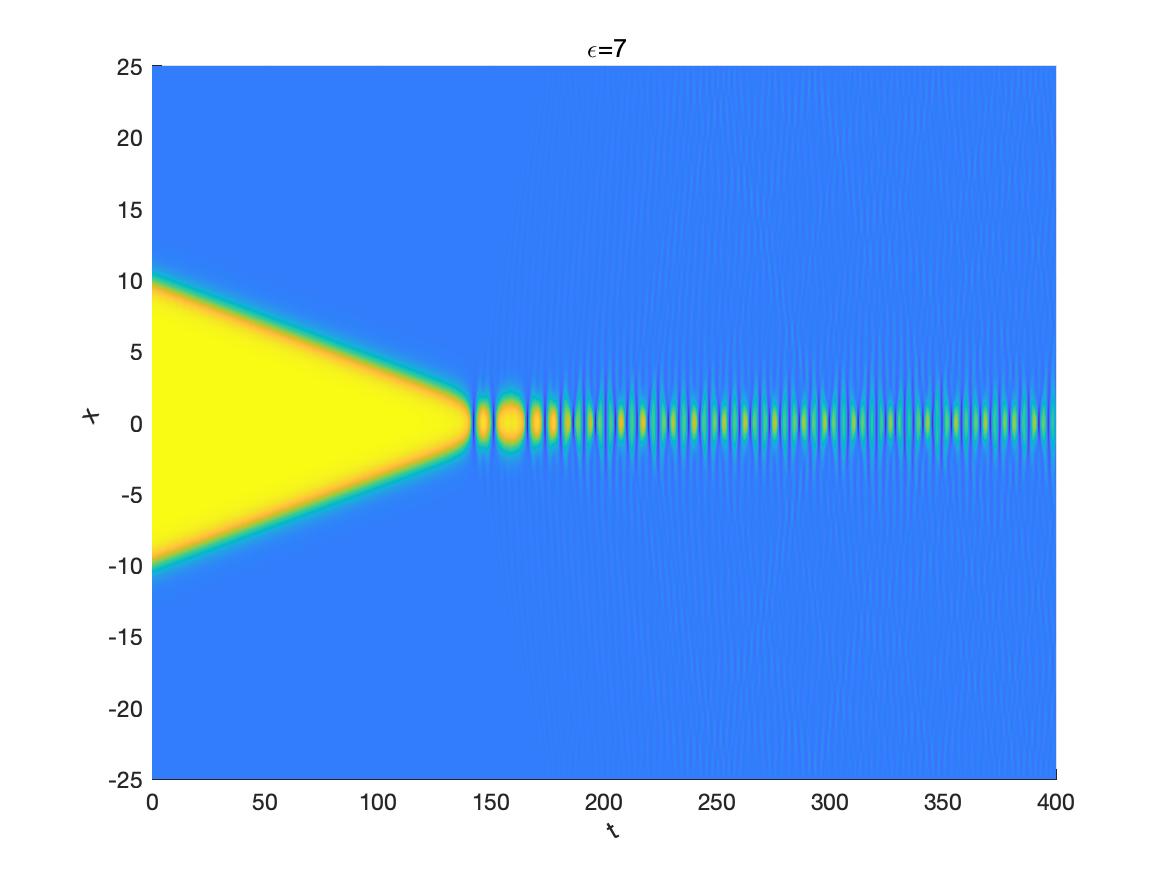} &
\hspace{-1em}
\includegraphics[width=.224\textwidth]{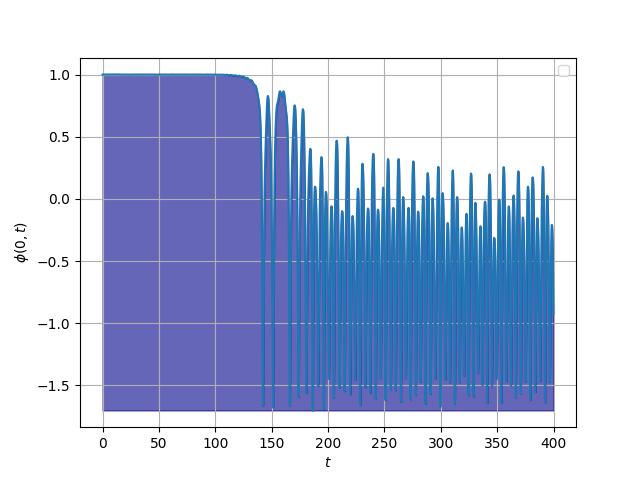} \\
\includegraphics[width=.224\textwidth]{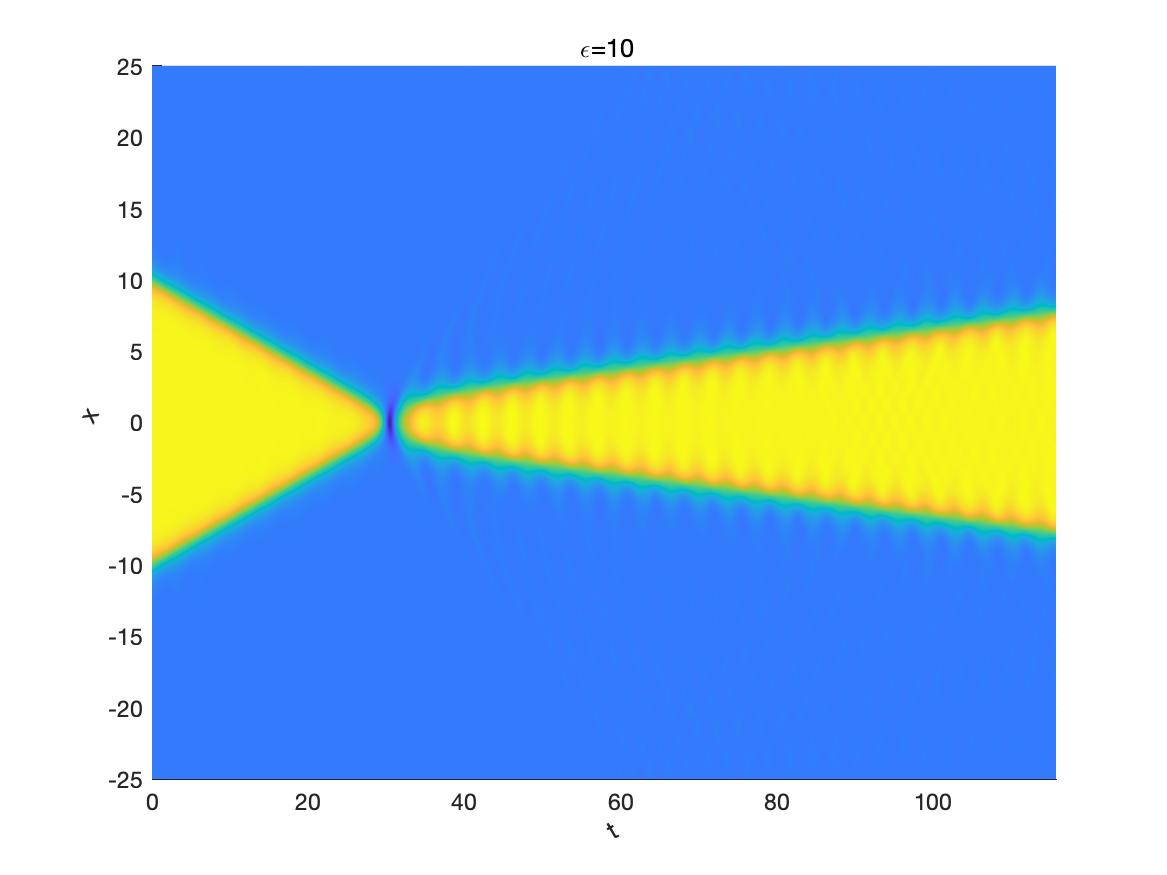} &
\hspace{-1em}
\includegraphics[width=.224\textwidth]{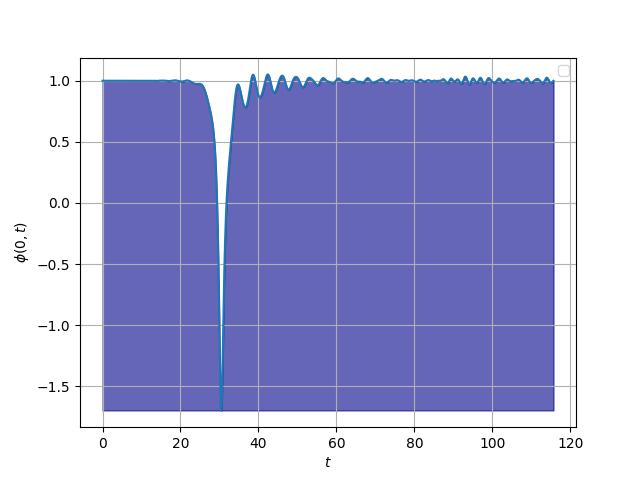} &
\hspace{-1em}
\includegraphics[width=.224\textwidth]{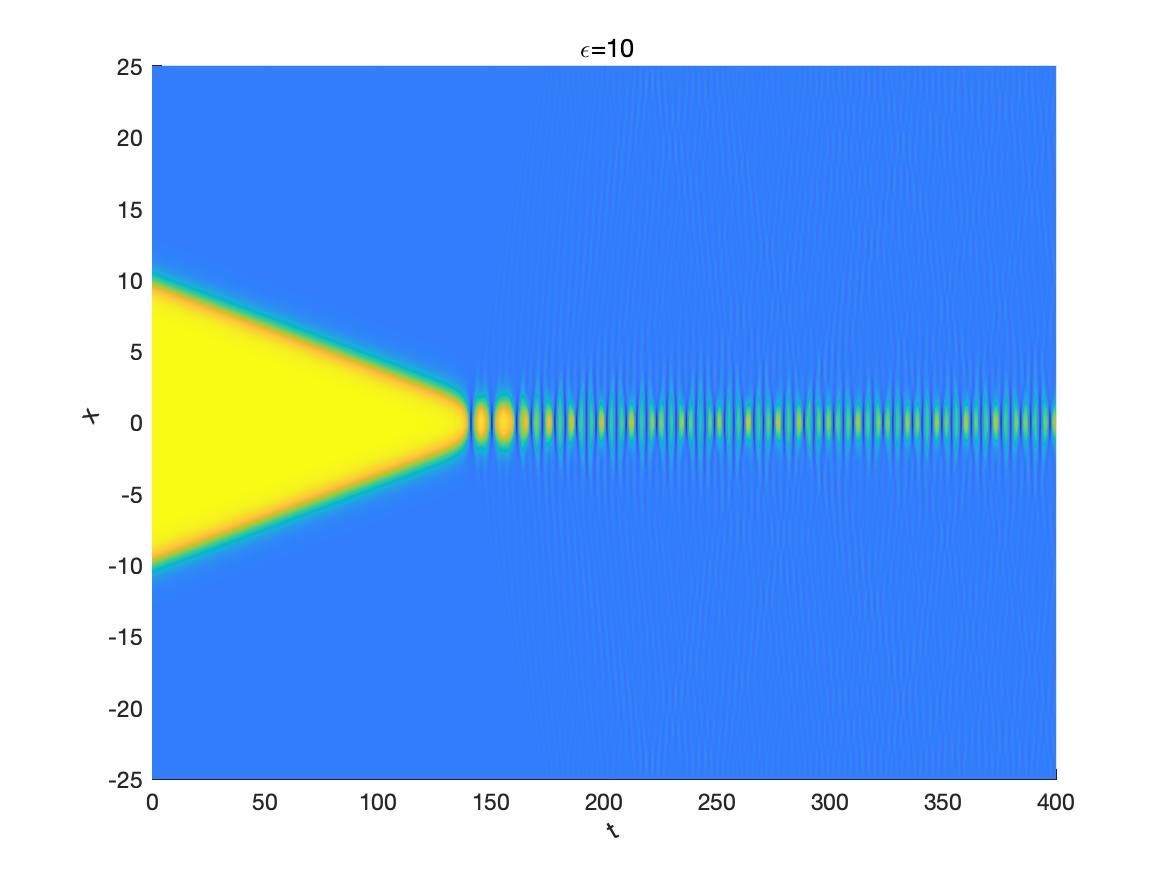} &
\hspace{-1em}
\includegraphics[width=.224\textwidth]{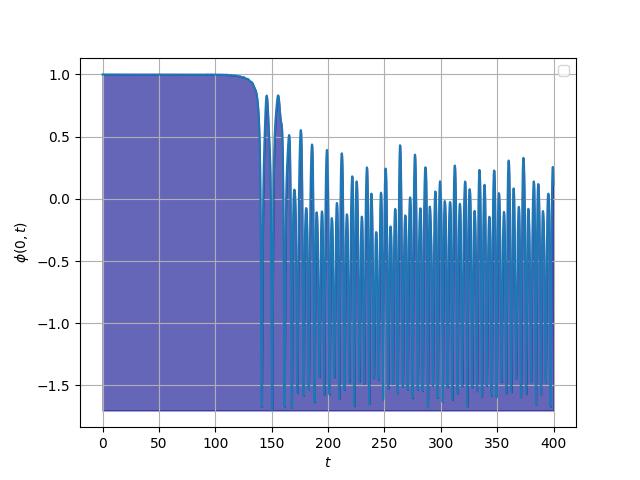} \\
\includegraphics[width=.224\textwidth]{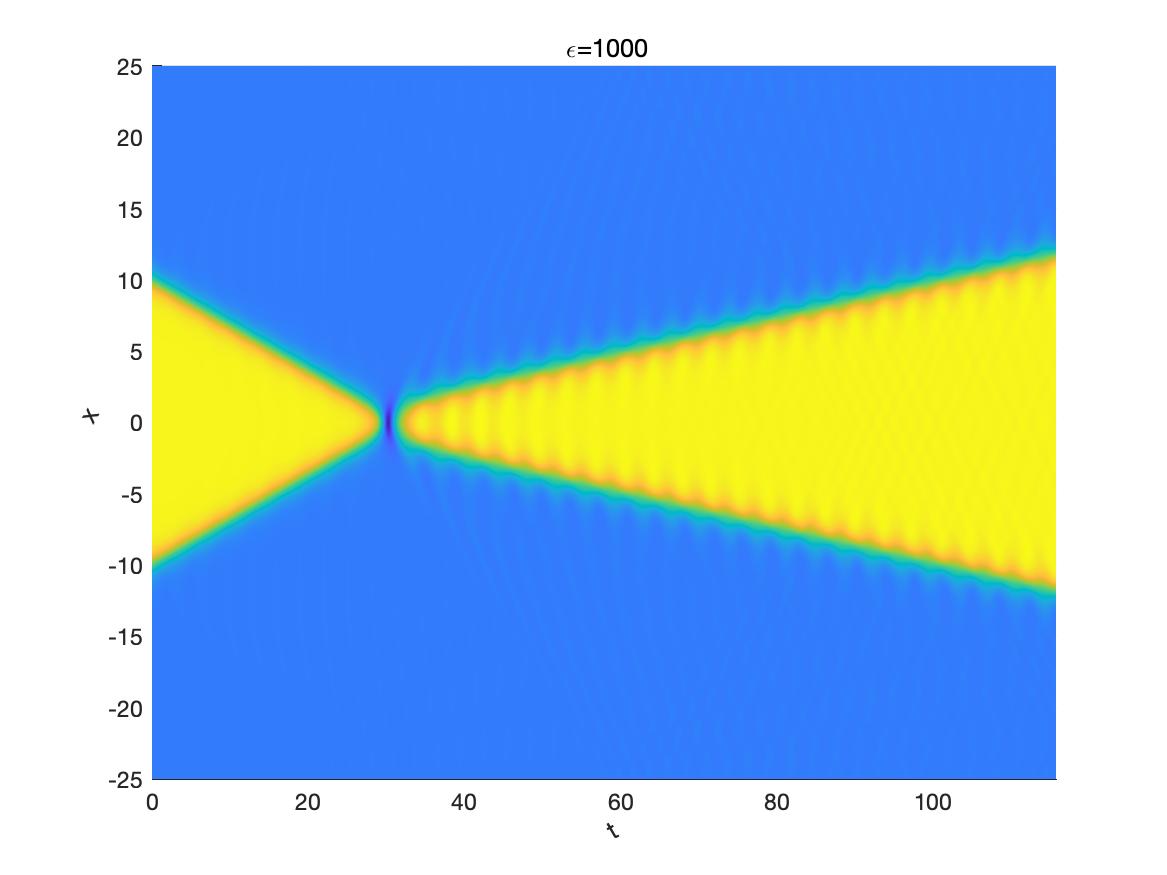} &
\hspace{-1em}
\includegraphics[width=.224\textwidth]{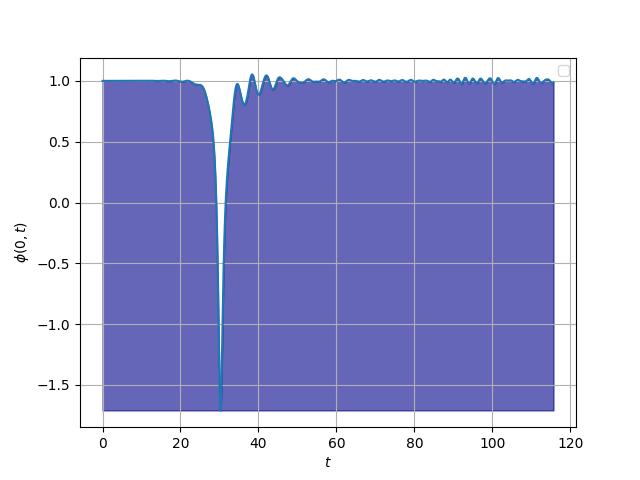} &
\hspace{-1em}
\includegraphics[width=.224\textwidth]{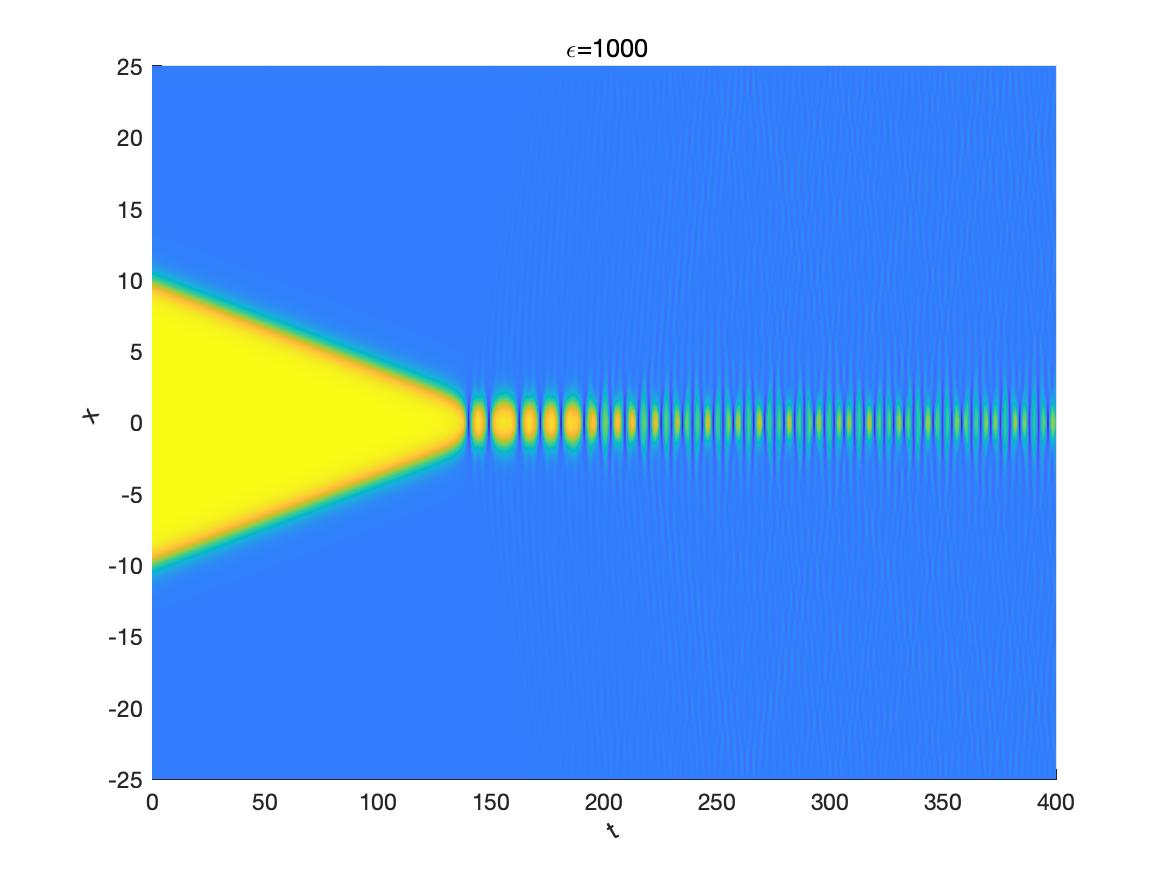} &
\hspace{-1em}
\includegraphics[width=.224\textwidth]{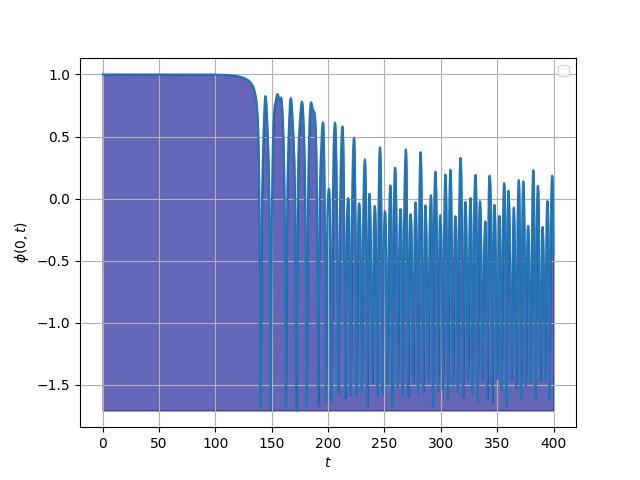} \\
\end{tabular}

\caption{The $\phi(x,t)$ and $\phi(0,t)$ images with  $\epsilon\in\{3,5,7,10,1000\}$ are shown from top to bottom panels.  The left two columns are the scatter state with $V_{\text{in}}=0.33$, and the right two columns are the bion state with $V_{\text{in}}=0.06$.\label{fig:c}}
\end{figure}

As shown on the left side of Figure~\ref{fig:c}, after a single collision, the kink and antikink separate with equal speed in opposite directions, which is called the scatter state. Besides this, there are cases involving two, three, four, or even more collisions. The number of collisions is referred to as a “$n$-bounce”.  For the right side of Figure~\ref{fig:c}, where $n$ goes to infinity, we usually term it the bion state~\cite{kudryavtsev1975solitonlike}.

\begin{figure}[htbp]
    \centering
    \subfigure{
        \includegraphics[width=.44\textwidth]{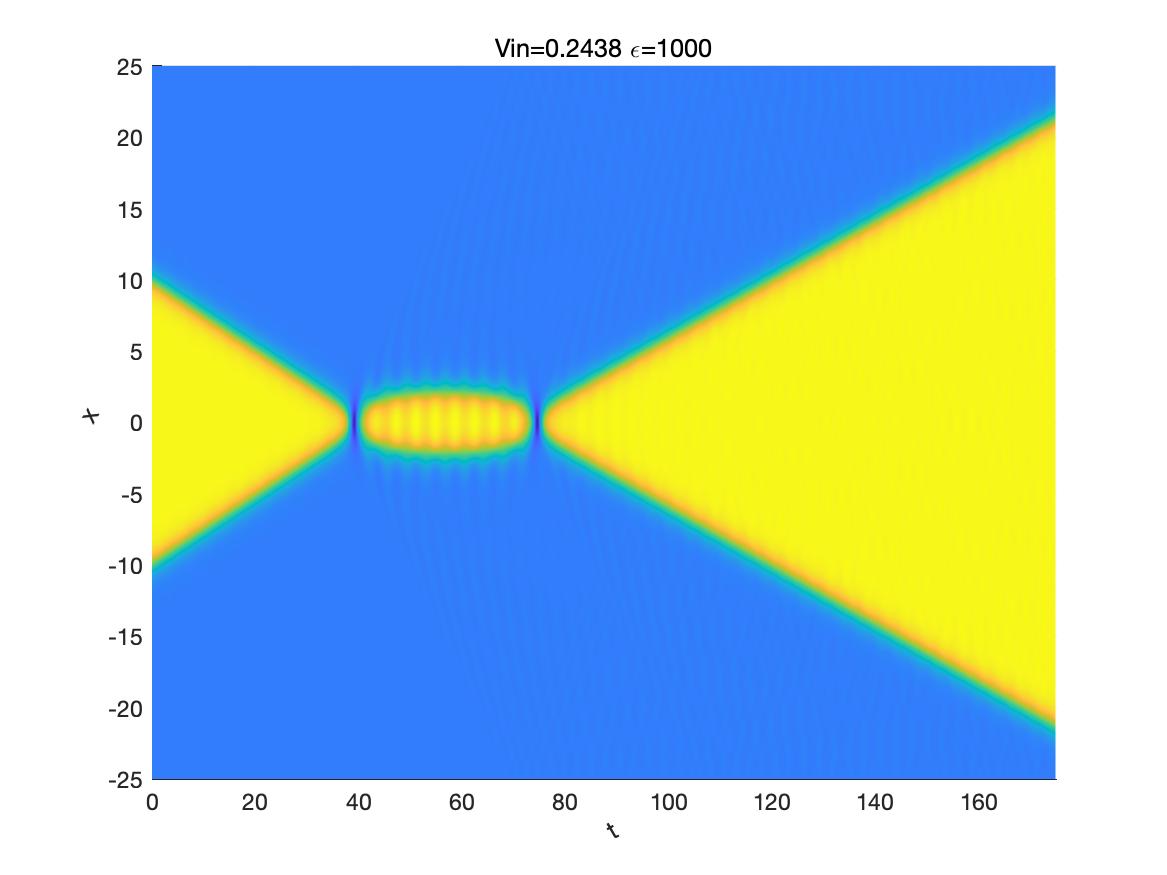}
        \label{fig:d-a}
    }
    \hspace{0.4em}
    \subfigure{
        \includegraphics[width=.44\textwidth]{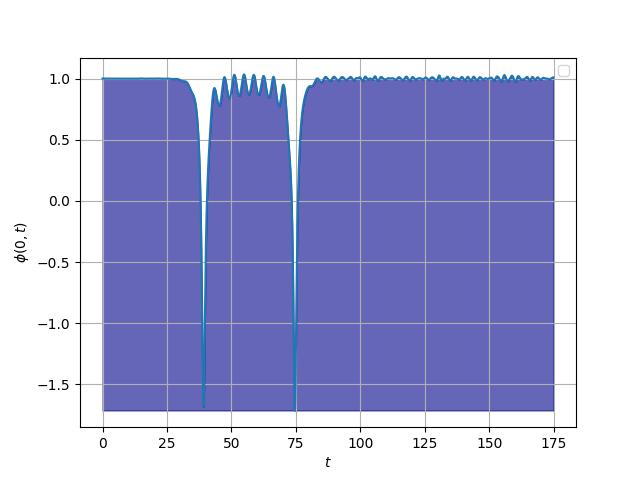}
        \label{fig:d-b}
    }
    \vspace{1em}
    \subfigure{
        \includegraphics[width=.44\textwidth]{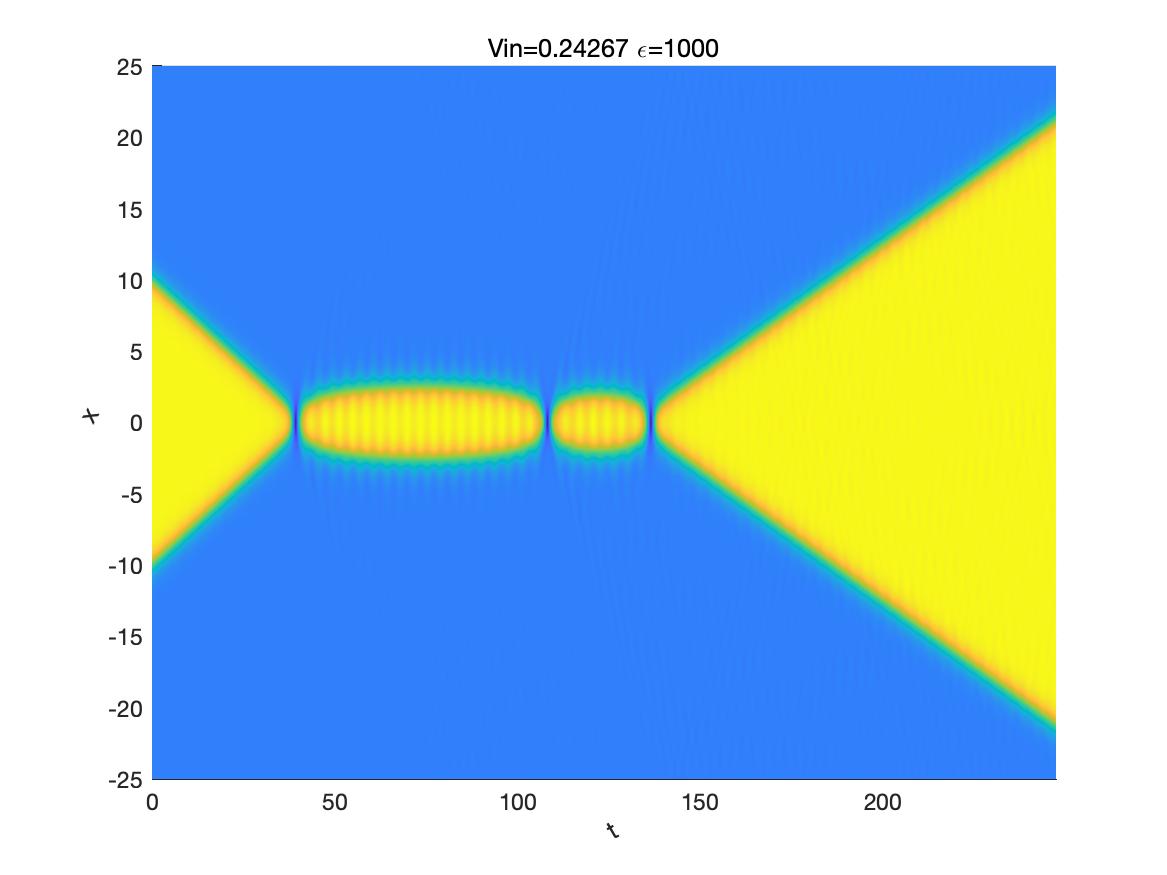}
        \label{fig:d-c}
    }
    \hspace{0.4em}
    \subfigure{
        \includegraphics[width=.44\textwidth]{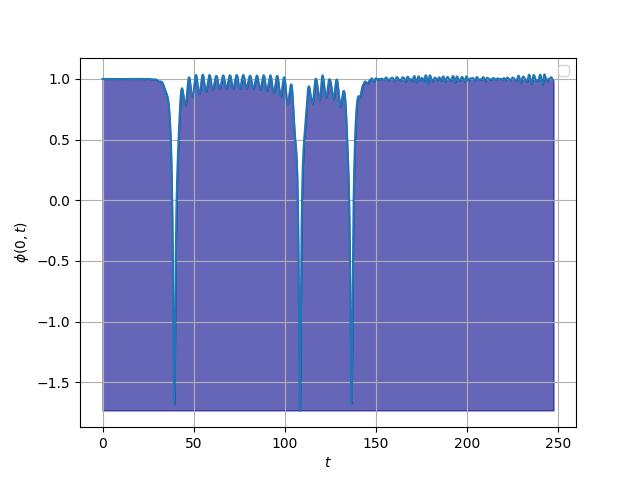}
        \label{fig:d-d}
    }
    \vspace{1em}
    \subfigure{
        \includegraphics[width=.44\textwidth]{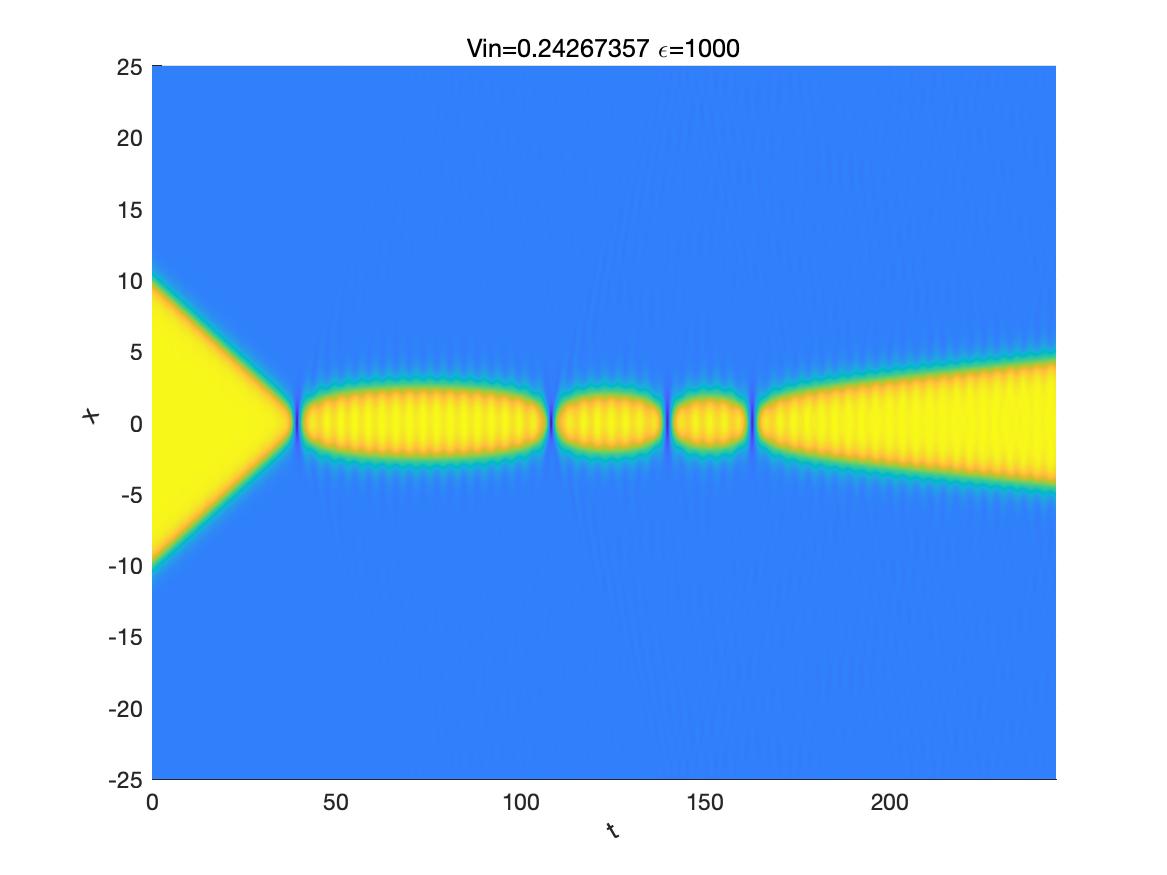}
        \label{fig:d-e}
    }
    \hspace{0.4em}
    \subfigure{
        \includegraphics[width=.44\textwidth]{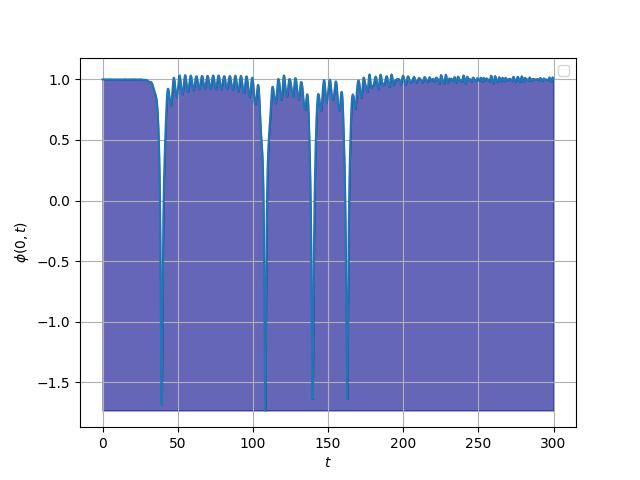}
        \label{fig:d-f}
    }
    \caption{\centering When $\epsilon=1000$, incident velocity $V_{\text{in}}$ values of 0.2438, 0.2426735, and 0.24267357 correspond to the 2-bounce state, 3-bounce state, and 4-bounce state, respectively.}
    \label{fig:d}
\end{figure}

The $n$-bounce states with finite $n$ occur only in certain intervals below the critical velocity $V_{\text{cr}}$. Using $\epsilon = 1000$ as an example, when $V_{\text{in}}$ = 0.2438, the numerical simulation results are shown in Figure~\ref{fig:d-a} and~\ref{fig:d-b}. This illustrates the collision, subsequent finite separation, re-collision, and eventual scatter of the kink and antikink, which is referred to as the two-bounce state. Similarly, for $\epsilon$=1000, there are 3-bounce state and 4-bounce state, corresponding to $V_{\text{in}}$=0.2426735 and $V_{\text{in}}$=0.24267357, as shown in the Figure~\ref{fig:d}.

From these cases study, it is evident that both the parameter $\epsilon$ and the incident velocity $V_{\text{in}}$ can significantly affect the collision between the kink and antikink. In detail, 
different $V_{\text{in}}$ can lead to different states, while different $\epsilon$ lead to the similar state but different escaping velocity. However, we claim that there is a bias of parameter selection.

As shown in the next section, different $\epsilon$ can also lead to different states with the same $V_{\text{in}}$, and can also show the fractal structure.

\section{Fractal structure}
\label{sec:four}
Similar to the $\phi^4$ theory, the CL model exhibits a discontinuous transition between scatter and trapped states as the incident velocity varies continuously. However, unlike other theories, the CL model also demonstrates this discontinuous behavior with continuous changes in $\epsilon$. 
In this section, we will  illustrate in detail how  $\epsilon$ and $V_{\text{in}}$ affect the dynamics of the CL model.

\subsection{The Self-similarity of fractal structure between \texorpdfstring{$V_{\text{out}} - V_{\text{in}}$}.}

\begin{figure}[htbp]
    \begin{tabular}{cc}
    \centering
    \includegraphics[width=0.49\textwidth,height=4.5cm]{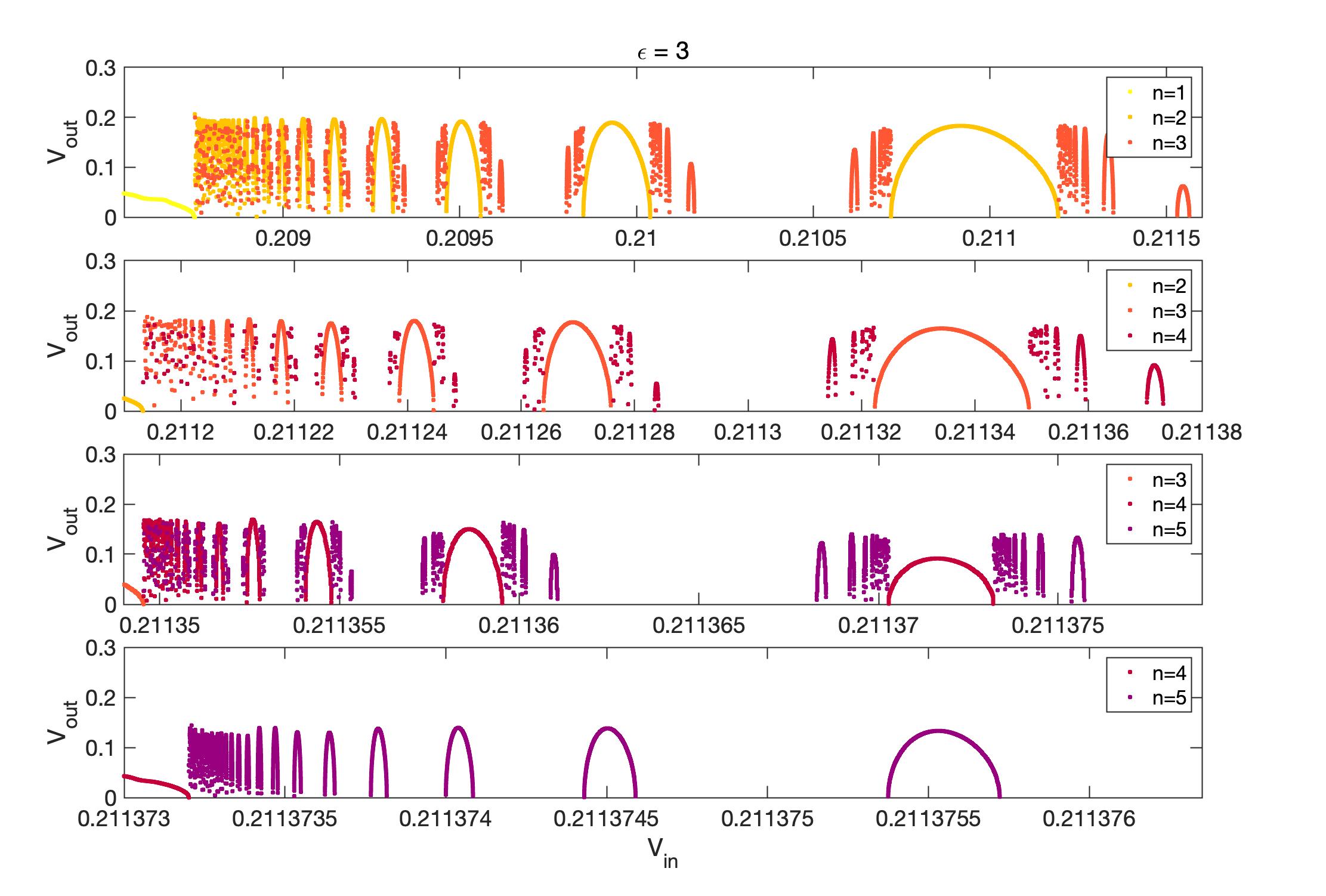}
        \label{fig:f-a}&\hspace{-1em}\includegraphics[width=0.49\textwidth,height=4.5cm]{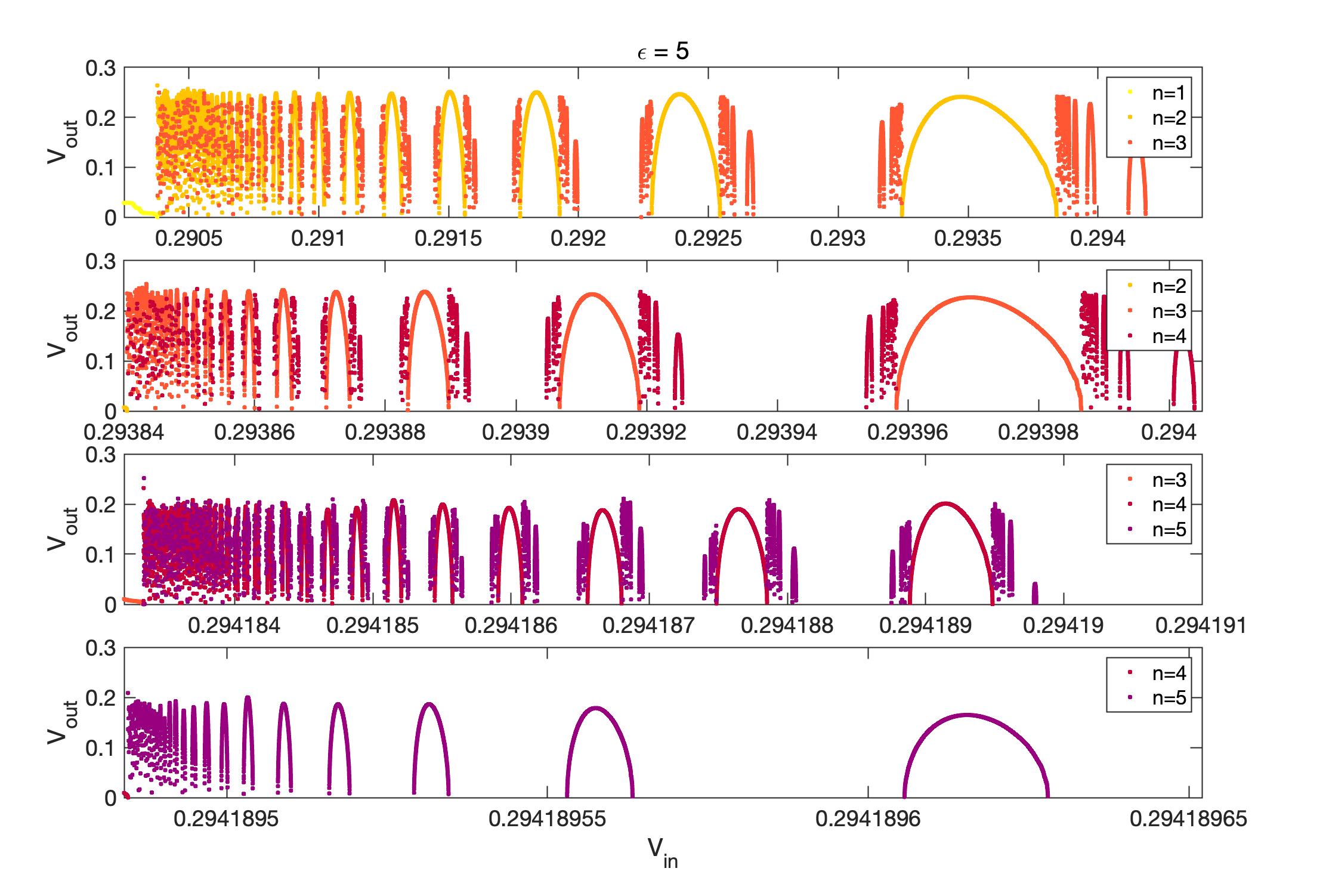}
        \label{fig:f-b}\\
    \includegraphics[width=0.49\textwidth,height=4.5cm]{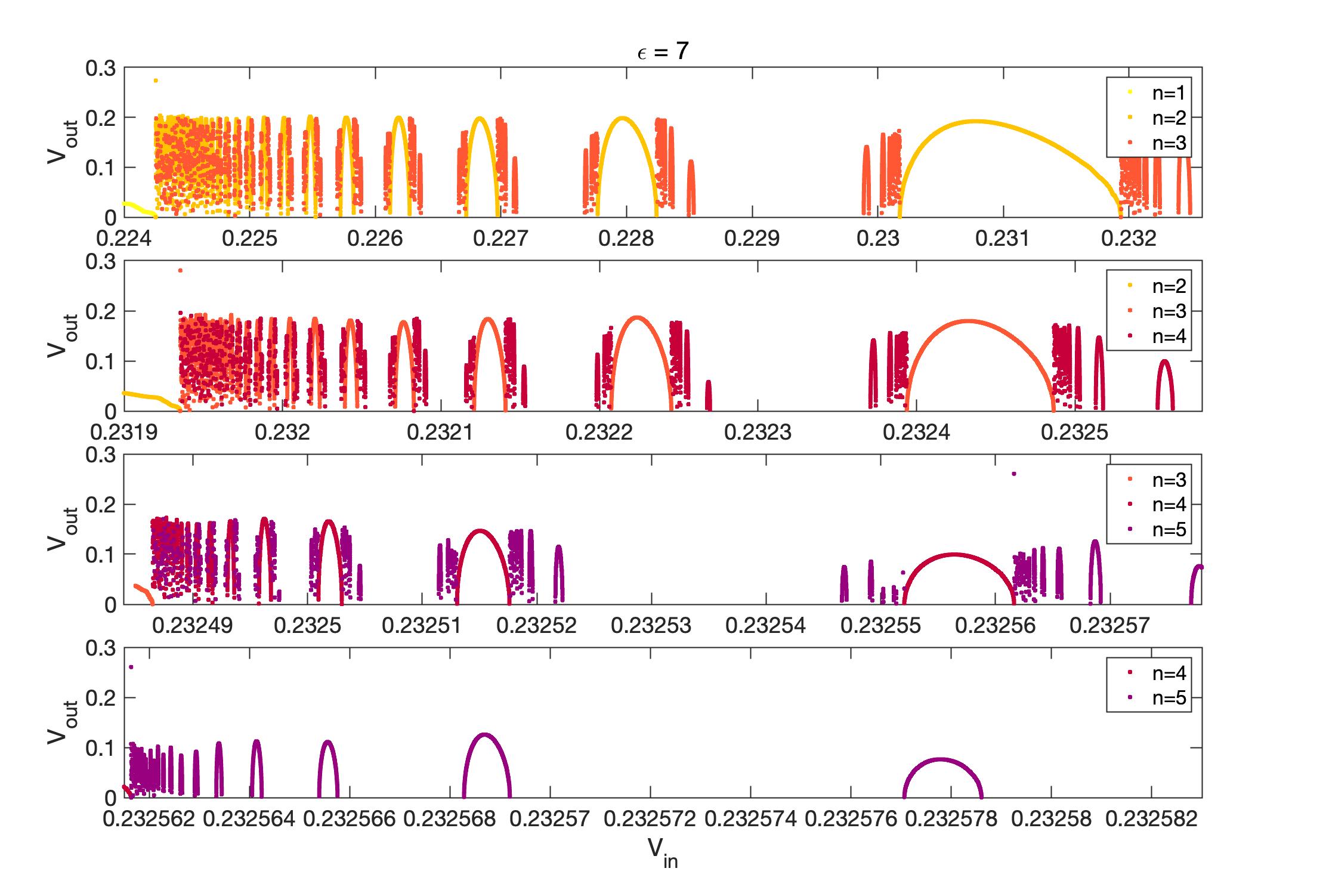}
            \label{fig:f-c}&\hspace{-1em}\includegraphics[width=0.49\textwidth,height=4.5cm]{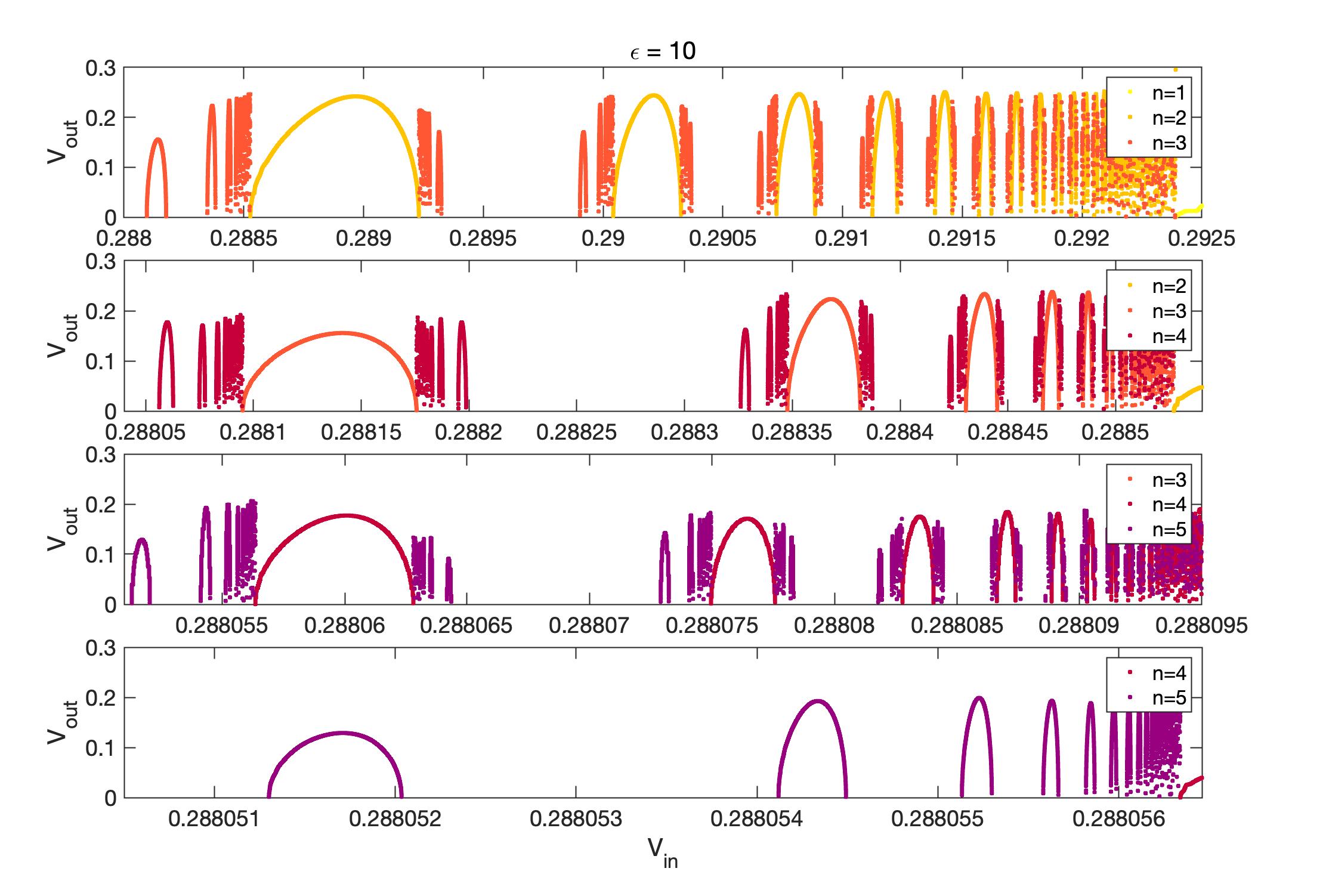}
        \label{fig:f-d}\\
    \end{tabular}
    \centering
    \begin{tabular}{c}
        \includegraphics[width=0.5\textwidth,height=4.5cm]{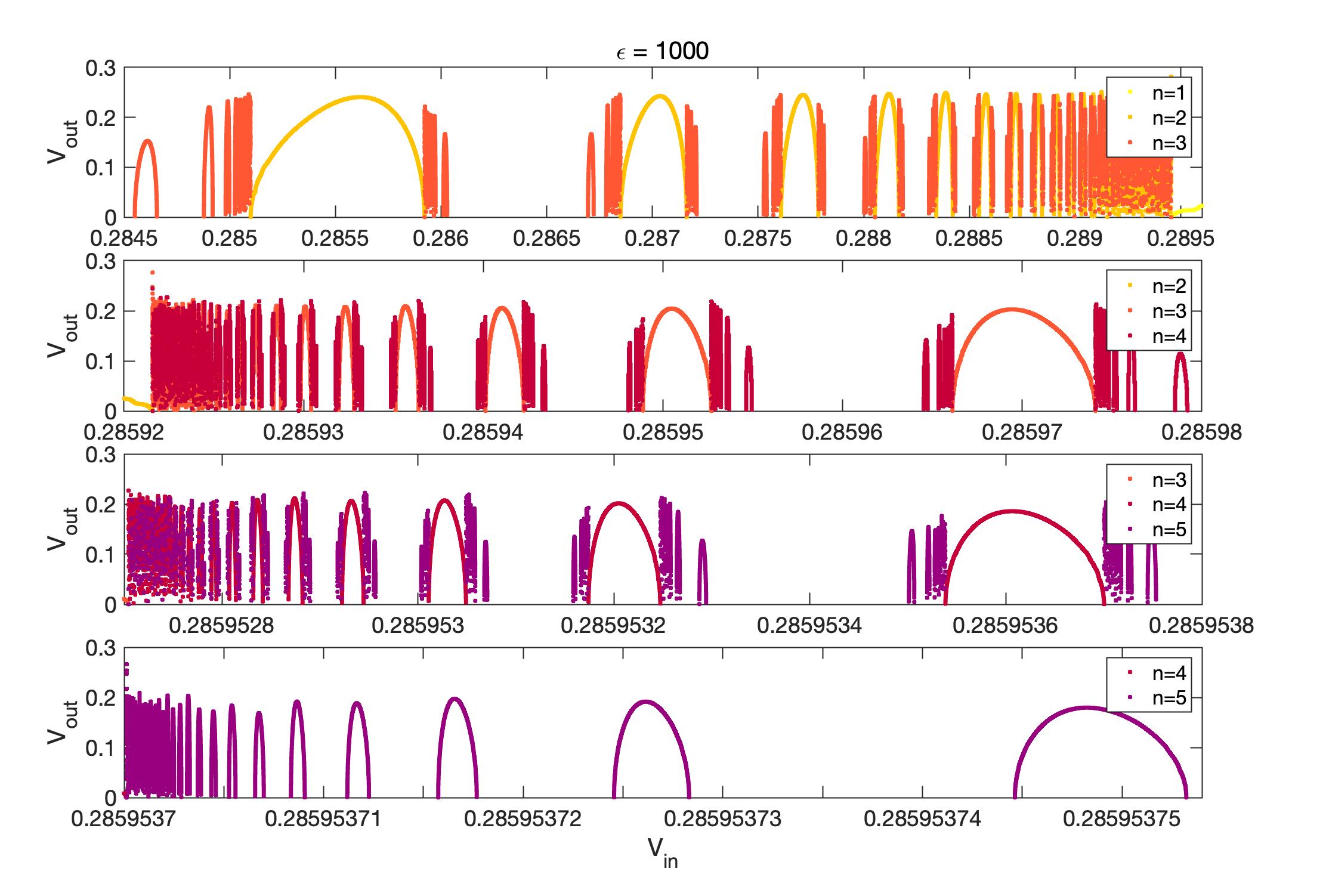}\\
        \label{fig:f-e}
    \end{tabular}
    \caption{\centering Fractal structure when $\epsilon \in \{3, 5, 7, 10, 1000\}$. The yellow, orange, red, and purple lines represent 2-bounce, 3-bounce, 4-bounce, and 5-bounce states, respectively.}
    \label{fig:f}
\end{figure}

First, we explain the numerical calculation  of $V_{\text{out}}$. We record the spatial positions of the kink and antikink from the last collision to $T_{\rm tru}$ or $T_{\rm tot}$, and the outgoing velocity $V_{\text{out}}$ is then obtained through linear fitting positions in the corresponding time range.
In Figure \ref{fig:f}, we present the plot of $V_{\text{out}}$ versus $V_{\text{in}}$  for  $\epsilon \in\{3,5,7,10,1000\}$.  In each panel of Figure~\ref{fig:f}, it is evident that the plots 
demonstrate notable self-similarity, which are of the fractal structures for the CL model.

We explain the $\epsilon = 3$ case (Figure~\ref{fig:f-a}) for illustration. From top to bottom, there are four $V_{\text{in}}$ intervals   (0.2086, 0.2116), (0.21119, 0.21119), (0.211349, 0.211379), and (0.211373, 0.211376), corresponding to four sets of increasing bounce states respectively.  The velocity interval for the 3-bounce state is a zoomed-in section of the rightmost part of the 2-bounce state. Similarly, the velocity interval for the 4-bounce state is a zoomed-in section of the rightmost part of the 3-bounce state, and the rightmost part of the 4-bounce state zooms in to form the velocity interval for the 5-bounce state.

It is worth noting that, as shown in Figure~\ref{fig:f}, the fractal structure patterns are similar for $\epsilon \in {3, 5, 7, 10}$. However, for $\epsilon = 1000$, the pattern of the 2-bounce state (top panel) differs from that of other three bounce states. This case is solely due to the selection bias of the velocity interval, and does not indicate a different fractal structure.

Finally, we observed that changes in \(\epsilon\) do not affect the self-similarity of the fractal structure. However, \(\epsilon\) significantly influences the critical velocity \(V_{\text{cr}}\), as detailed in Section~\ref{sec:fourthree}.

\subsection{The Fractal structure of \texorpdfstring{$V_{\text{out}}$}. versus \texorpdfstring{$\epsilon$}.}

In the previous part, we presented a kinetic analysis for $\epsilon \in \{3, 5, 7, 10, 1000\}$ and also show their $V_{\text{out}}$-$V_{\text{in}}$ fractal structure. We observe that $\epsilon$ can affect the dynamics of the collisions. However, further investigation is required to elucidate the detailed evolution of collision behavior as a function of $\epsilon$.

\begin{figure}[htbp]
    \centering
    \subfigure{
        \includegraphics[width=.8\textwidth,height=6cm]{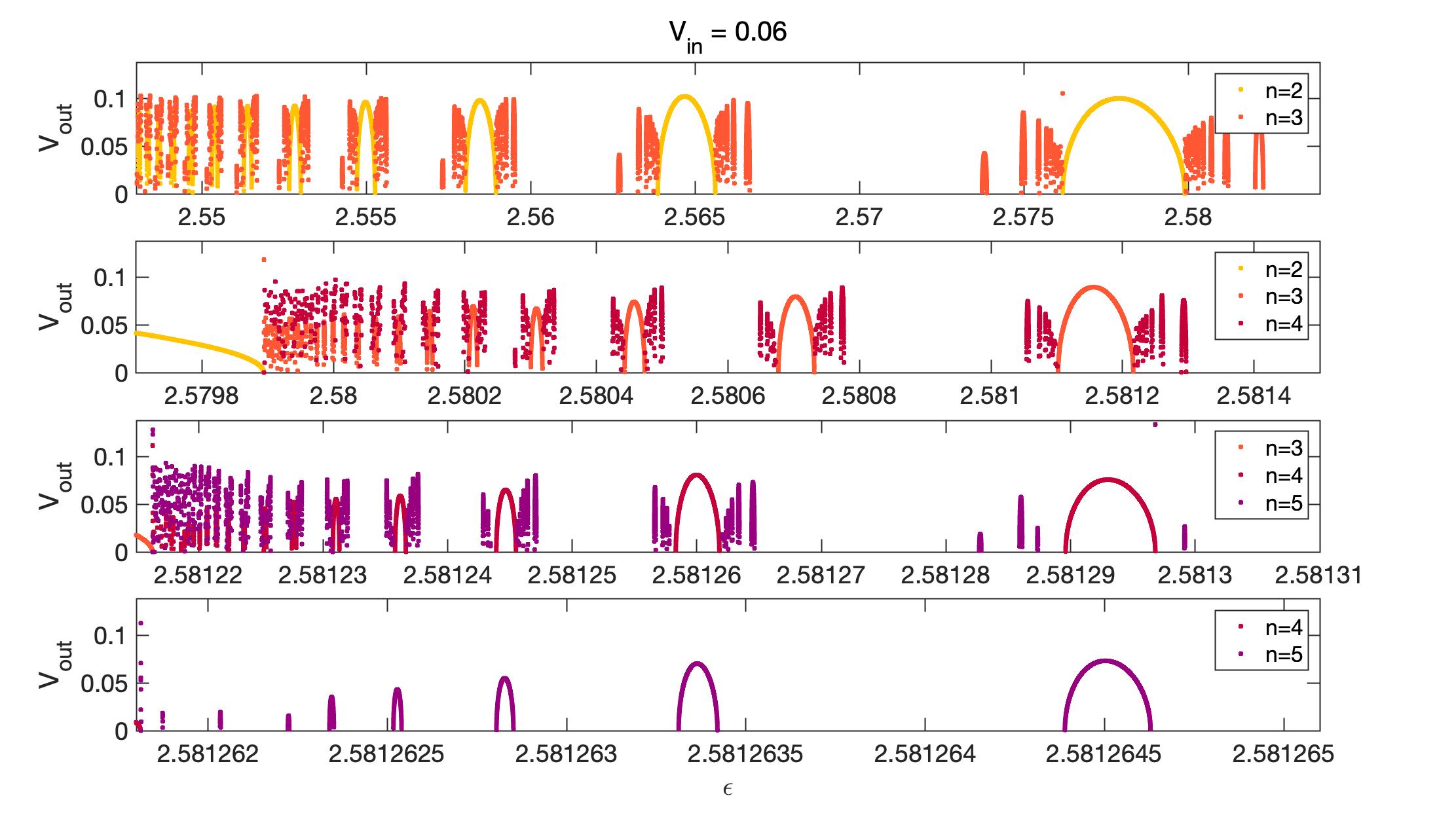}
        \label{fig:g-a}
    }
    \vspace{1em}
    \subfigure{
        \includegraphics[width=.8\textwidth,height=6cm]{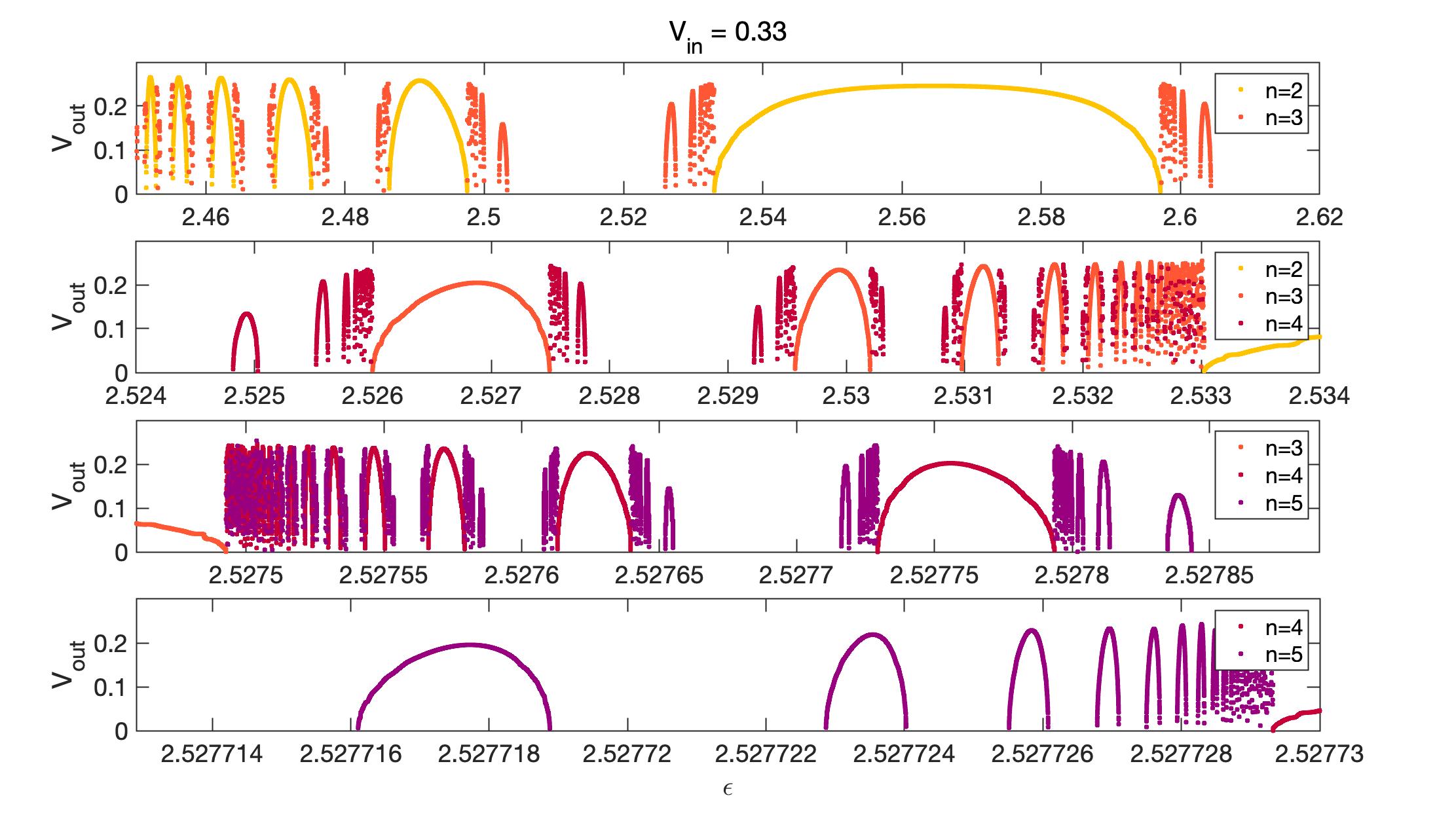}
        \label{fig:g-b}
    }
    \caption{The plot of $V_{\text{out}}$ versus $\epsilon$. The  above and below panels correspond to $V_{\text{in}}$=0.06 and $V_{\text{in}}$=0.33, respectively.}
    \label{fig:g}
\end{figure}

Given the incident velocity $V_{\text{in}}$, we illustrate how the $V_{\text{out}}$  changes when the parameter $\epsilon$ is continuously tuned. 
In this study, we have chosen two incident velocities $V_{\text{in}}=0.06$ and $V_{\text{in}}=0.33$, which have been illustrated as the example in Section~\ref{sec:three}.  For the case of $V_{\text{in}}=0.06$ (Figure~\ref{fig:g-a}), it is clearly observed that the velocity intervals for the four panels are (2.548, 2.584), (2.5797, 2.5815), (2.581215, 2.581311), and (2.5812618, 2.5812651), respectively. Each subsequent bounce state is a zoomed-in section of the rightmost part of the previous bounce state.

For the incident velocity $V_{\text{in}}=0.33$, the chosen velocity intervals still maintain a high degree of self-similarity. Thus, we claim that there is a fractal structure in the plot $V_{\text{out}}$ versus $\epsilon$ for the CL model. 

The emergent fractal structure by tuning $\epsilon$ may be due to that the potential attend the energy transition process among the transitional, shape mode and potential. In the collective coordinate method, the kink and antikink are quasi-particles, they will pass the potential well, and share energy to the shape mode. Since the potential is $\epsilon$ dependent, the collision process will be effectively involved. However, we have no quantitative method to show the CCM solution of the CL model, which will further help to understand the emergence of the fractal structure. 


\subsection{t-\texorpdfstring{$V_{\text{in}}$}.}
\label{sec:fourthree}

\begin{figure}[htbp]
    \centering
    \subfigure[$\epsilon = 3$]{
        \includegraphics[width=0.48\textwidth]{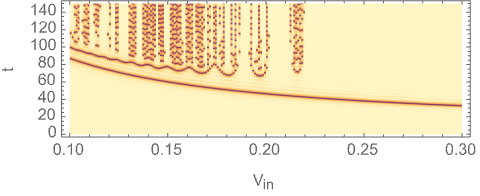}
        \label{fig:h-a}
        }
    \hspace{-1em}
    \subfigure[$\epsilon = 5$]{
        \includegraphics[width=0.48\textwidth]{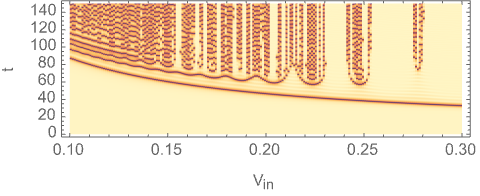}
        \label{fig:h-b}
        }
    \vspace{1mm}
    \subfigure[$\epsilon = 7$]{
        \includegraphics[width=.48\textwidth]{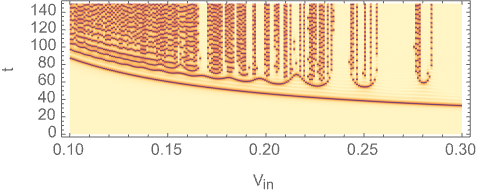}
        \label{fig:h-c}
        }
    \hspace{-1em}
    \subfigure[$\epsilon = 10$]{
        \includegraphics[width=.48\textwidth]{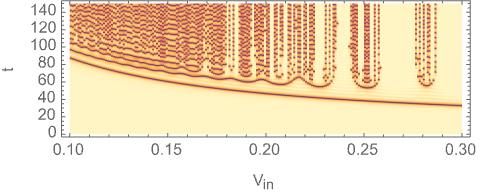}
        \label{fig:h-d}
        }
    \vspace{1mm}
    \subfigure[$\epsilon = 1000$]{
        \includegraphics[width=0.48\textwidth]{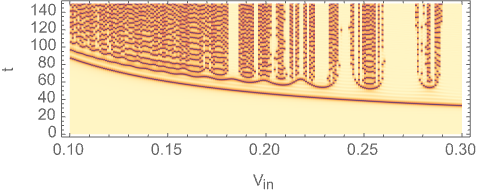}
        \label{fig:h-e}
        }
    
    \caption{\centering The central field values of $\phi(0, t)$ for $\epsilon$ = 3,5,7,10,1000 are shown with time and incident velocity respectively.}
    \label{fig:h}
\end{figure}

In Figure~\ref{fig:h}, we plot the center field value $\phi(0,t)$ as a function of time and incident velocity, illustrating the cases for $\epsilon = 3, 5, 7, 10, 1000$ in panels (a), (b), (c), (d), and (e), respectively. It is evident that as $\epsilon$ increases, the critical velocity $V_{cr}$ also increases, though at a progressively slower rate, eventually approaching a constant value when $\epsilon$ becomes sufficiently large, as observed by C. Adam in ~\cite{adam2023relativistic}.

For $\epsilon = 3$ (as shown in Figure~\ref{fig:h-a}), we find two missing bounce windows (MBW), around at $V_{\text{in}} = 0.19$ and $V_{\text{in}} = 0.21$. The MBWs shift towards higher velocities as $\epsilon$ increases. When $\epsilon = 5$ and $\epsilon = 7$ (as shown in Figure~\ref{fig:h-b} and~\ref{fig:h-c}), the two MBWs are observed at $V_{\text{in}} = 0.24$ and $V_{\text{in}} = 0.27$.  And as $\epsilon$ continues to increase, the positions of MBW stabilize.

\subsection{Hausdorff dimension}

In the previous part, we discussed the fractal structures of $V_{\text{out}}-V_{\text{in}}$ and $V_{\text{out}}-\epsilon$. In order to understand  the fractal structure quantitatively, we calculate the dimension of fractal structure.

There are two main definitions of fractal dimension: the Hausdorff dimension $D_H$~\cite{fernandez2012fractal} and the Box-counting dimension $D_B$~\cite{gorski2011accuracy}. The Hausdorff dimension has advantage in analytical properties, but is difficult to calculate. Generally, we choose the Box-counting dimension for practical computation, and it can be shown that the box dimension can serve as a substitute for the Hausdorff dimension in many cases. 

\begin{figure}[htbp]
    \centering
    \includegraphics[width=0.7\textwidth]{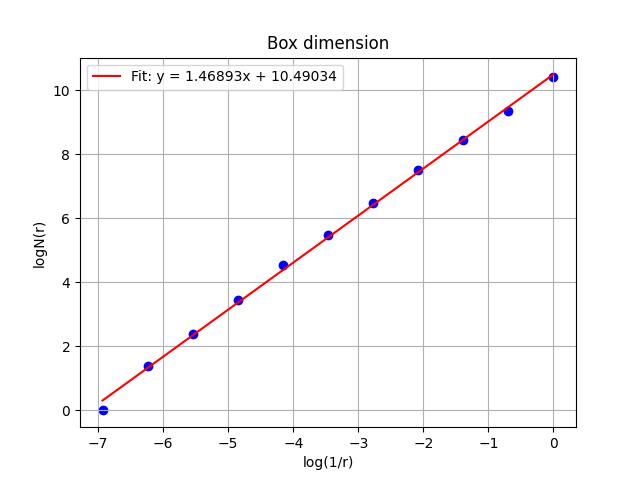}
    \caption{\centering The plot of $\log N(r)-\log(\frac{1}{r})$. Its slope is the fractal dimension of the graph.}
    \label{fig:i}
\end{figure}

To calculate the Box-counting dimension, the image is processed to become a binary black and white image matrix  with black representing 0 and white representing 1. By expanding the matrix into an operationally convenient $k \times k$ matrix, where $k$ represents the maximum number of rows and columns in the array.  The size of the matrix is set to be 1, and then dividing it by a specific dimension $r$, the matrix is partitioned into $(\frac{1}{r} \times \frac{1}{r})$ smaller grids. The values in the binary matrix are reversed that grids containing the image are set to 1, while those without the image are set to 0. Then, by summing the values of each grid in the matrix, the total number of grids containing the image, denoted as $N(r)$, is obtained. Taking  different  $r=(2^0,2^1,... ,2^p)/2^p$, where $2^p$ is greater than or equal to $k$, and $2^{(p-1)}$ is less than $k$,  a set of $N(r)$ can be obtained to plot the $\log N(r)$ versus $\log(\frac{1}{r})$ images. The Box-counting dimension is defined as 
\begin{equation}
D_B = \lim_{r \to \infty} \frac{\log N(r)}{\log (\frac{1}{r})}.
\end{equation}
The slope of the linear fitting between $\log N(r)$ and $\log (\frac{1}{r})$  represents the Box-counting dimension $D_B$, as illustrated in Figure~\ref{fig:i}.

We compute the box dimension for the fractal structures presented in Sections 4.1 and 4.2, and all results are collected in Table~\ref{tab:1} and Table~\ref{tab:2}.

\begin{table}[htbp]
\centering
\begin{tabularx}{0.8\textwidth}{>{\raggedright\arraybackslash}X @{\hskip 1cm} >{\raggedleft\arraybackslash}X}
\hline
$V_{in}-V_{out}$&Box-dimension\\
\hline
0.208-0.215\footnotemark[1]&1.46010471127\\
0.21119-0.2114\footnotemark[1]&1.35988119125\\
0.211349-0.2114\footnotemark[1]&1.35988119625\\
0.211373-0.211376\footnotemark[1]&1.42757608141\\
\hline
0.2889-0.2945\footnotemark[2]&1.52627136107\\
0.29384-0.29419\footnotemark[2]&1.48211054884\\
0.2941832-0.29419\footnotemark[2]&1.55483390326\\
0.294189484-0.294189805\footnotemark[2]&1.46511657077\\
\hline
0.224-0.2328\footnotemark[3]&1.55088184693\\
0.2318-0.2326\footnotemark[3]&1.48134850030\\
0.232485-0.23258\footnotemark[3]&1.44555481112\\
0.2325615-0.23258\footnotemark[3]&1.42813726066\\
\hline
0.287-0.2925\footnotemark[4]&1.57548884511\\
0.28805-0.28855\footnotemark[4]&1.47796118684\\
0.28805-0.288095\footnotemark[4]&1.51763753302\\
0.2880512-0.28805646\footnotemark[4]&1.46150279699\\
\hline
0.2845-0.2896\footnotemark[5]&1.66776038939\\
0.28592-0.285983\footnotemark[5]&1.46892587647\\
0.2859527-0.2859538\footnotemark[5]&1.56780778167\\
0.2859537-0.285953754\footnotemark[5]&1.57028337778\\
\hline

\end{tabularx}
\caption{\centering The fractal box dimension for $\epsilon$=3, 5, 7, 10, 1000 are listed  from top to bottom panel, respectively\label{tab:1}}
\footnotetext[1]{The $V_\text{in}$ range when $\epsilon$ is 3}
\footnotetext[2]{The $V_\text{in}$ range when $\epsilon$ is 5}
\footnotetext[3]{The $V_\text{in}$ range when $\epsilon$ is 7}
\footnotetext[4]{The $V_\text{in}$ range when $\epsilon$ is 10}
\footnotetext[5]{The $V_\text{in}$ range when $\epsilon$ is 1000}

\end{table}

\begin{table}[htbp]
\centering
\begin{tabularx}{0.8\textwidth}{>{\raggedright\arraybackslash}X @{\hskip 1cm} >{\raggedleft\arraybackslash}X}
\hline
$\epsilon$&Box-dimension\\
\hline
2.52-2.7\footnotemark[1]&1.38075775785\\
2.524-2.5336\footnotemark[1]&1.50775438500\\
2.52747-2.52789\footnotemark[1]&1.47740766104\\
2.527715-2.52773\footnotemark[1]&1.50310681105\\
\hline
2.548-2.5825\footnotemark[2]&1.53128385062\\
2.5798-2.5813\footnotemark[2]&1.42211584283\\
2.581215-2.5813\footnotemark[2]&1.47696860664\\
2.5812618-2.5812648\footnotemark[2]&1.33735408084\\
\hline

\end{tabularx}

\caption{\centering The fractal box dimension\label{tab:2} of $\epsilon-V_{\text{out}}$ for $V_{\text{in}} =0.33$ and $0.06$ are listed. }
\footnotetext[1]{The $\epsilon$ range when $V_\text{in}$ is 0.33}
\footnotetext[2]{The $\epsilon$ range when $V_\text{in}$ is 0.06}
\end{table}

\begin{figure}[htbp]
    \centering
    \includegraphics[width=0.7\textwidth]{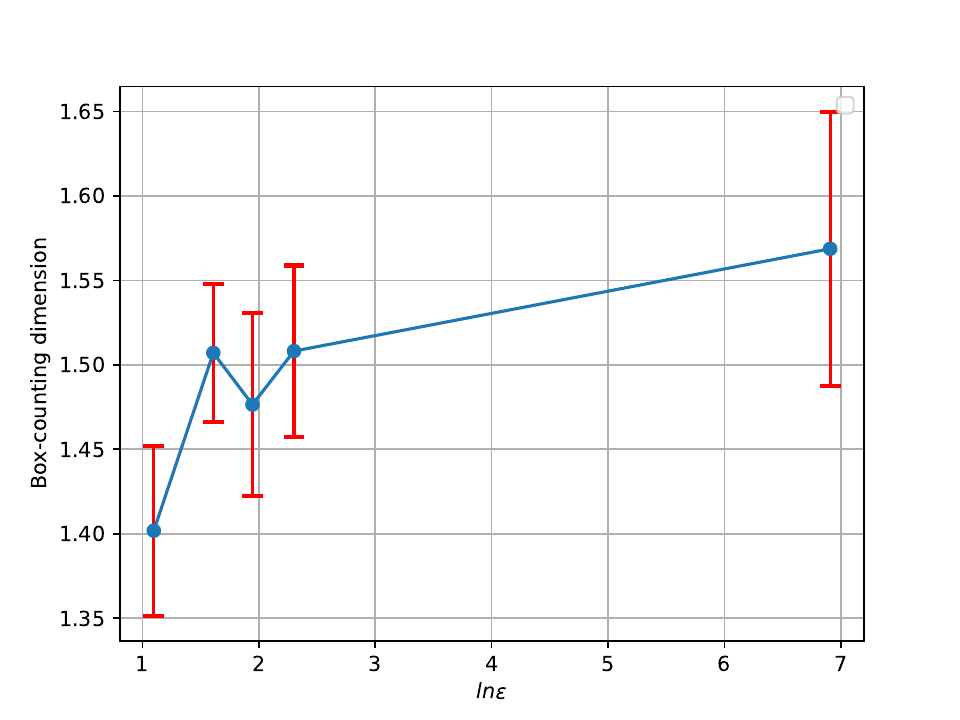}
    \caption{\centering The plot of the Box-counting dimension versus $\ln\epsilon$.}
    \label{fig:j}
\end{figure}

The fractal Box-counting dimension of the CL model ranges between 1.36 and 1.66, with similar fractal dimensions for given $\epsilon$. Furthermore, as shown in Figure~\ref{fig:j}, the average fractal Box-counting dimension increases  with  $\ln \epsilon$. When \(\epsilon\) becomes sufficiently large (mostly close to $\phi^4$ theory), the fractal dimension of the CL model is around 1.58, which is the Hausdorff dimension of the Sierpinski triangle, i.e., $\log 3 / \log 2 \sim 1.58$. The theoretical explanation for this dimension is still missing.

\section{Conclusion}\label{sec12}

Our work focuses on investigating the fractal structure of the Christ-Lee (CL) model. Initially, we demonstrate how the potential of the CL model changes with different parameters of \(\epsilon\), revealing its unique properties that bridge the \(\phi^4\) and \(\phi^6\) theories. We present numerical results for different values of \(\epsilon\) and  $V_{\text{in}}$, showing the emergence of bion states, scatter states, and $n$-bounce states. Additionally, we study  the fractal structure of the CL model to explore the dynamics of soliton interactions. Our analysis reveals a high degree of similarity in the \(V_{\text{out}} \) versus \(V_{\text{in}}\) behavior across different \(\epsilon\) values. We observe that changes in \(\epsilon\) significantly influence the critical velocity \(V_{\text{cr}}\) in the \(t - V_{\text{in}}\) plots. Given the presence and impact of the parameter \(\epsilon\) in the CL model, we also find a similar self-similarity in \(V_{\text{out}}\) versus \(\epsilon\), which has not been reported in historical literature. Finally, we employ the Box-counting dimension as a measure of the self-similarity of the fractal structures in \(V_{\text{out}}\) versus \(V_{\text{in}}\) and \(V_{\text{out}}\) versus \(\epsilon\). Notably, we observe that, as \(\epsilon\) increases, the Box-counting dimension also increases and approaches the Hausdorff dimension of 1.58, characteristic of the Sierpinski triangle.

\backmatter




\bibliography{sn-article}


\begin{thebibliography}{14}
\ifx \bisbn   \undefined \def \bisbn  #1{ISBN #1}\fi
\ifx \binits  \undefined \def \binits#1{#1}\fi
\ifx \bauthor  \undefined \def \bauthor#1{#1}\fi
\ifx \batitle  \undefined \def \batitle#1{#1}\fi
\ifx \bjtitle  \undefined \def \bjtitle#1{#1}\fi
\ifx \bvolume  \undefined \def \bvolume#1{\textbf{#1}}\fi
\ifx \byear  \undefined \def \byear#1{#1}\fi
\ifx \bissue  \undefined \def \bissue#1{#1}\fi
\ifx \bfpage  \undefined \def \bfpage#1{#1}\fi
\ifx \blpage  \undefined \def \blpage #1{#1}\fi
\ifx \burl  \undefined \def \burl#1{\textsf{#1}}\fi
\ifx \doiurl  \undefined \def \doiurl#1{\url{https://doi.org/#1}}\fi
\ifx \betal  \undefined \def \betal{\textit{et al.}}\fi
\ifx \binstitute  \undefined \def \binstitute#1{#1}\fi
\ifx \binstitutionaled  \undefined \def \binstitutionaled#1{#1}\fi
\ifx \bctitle  \undefined \def \bctitle#1{#1}\fi
\ifx \beditor  \undefined \def \beditor#1{#1}\fi
\ifx \bpublisher  \undefined \def \bpublisher#1{#1}\fi
\ifx \bbtitle  \undefined \def \bbtitle#1{#1}\fi
\ifx \bedition  \undefined \def \bedition#1{#1}\fi
\ifx \bseriesno  \undefined \def \bseriesno#1{#1}\fi
\ifx \blocation  \undefined \def \blocation#1{#1}\fi
\ifx \bsertitle  \undefined \def \bsertitle#1{#1}\fi
\ifx \bsnm \undefined \def \bsnm#1{#1}\fi
\ifx \bsuffix \undefined \def \bsuffix#1{#1}\fi
\ifx \bparticle \undefined \def \bparticle#1{#1}\fi
\ifx \barticle \undefined \def \barticle#1{#1}\fi
\bibcommenthead
\ifx \bconfdate \undefined \def \bconfdate #1{#1}\fi
\ifx \botherref \undefined \def \botherref #1{#1}\fi
\ifx \url \undefined \def \url#1{\textsf{#1}}\fi
\ifx \bchapter \undefined \def \bchapter#1{#1}\fi
\ifx \bbook \undefined \def \bbook#1{#1}\fi
\ifx \bcomment \undefined \def \bcomment#1{#1}\fi
\ifx \oauthor \undefined \def \oauthor#1{#1}\fi
\ifx \citeauthoryear \undefined \def \citeauthoryear#1{#1}\fi
\ifx \endbibitem  \undefined \def \endbibitem {}\fi
\ifx \bconflocation  \undefined \def \bconflocation#1{#1}\fi
\ifx \arxivurl  \undefined \def \arxivurl#1{\textsf{#1}}\fi
\csname PreBibitemsHook\endcsname

\bibitem[\protect\citeauthoryear{Dorey et~al.}{2011}]{dorey2011kink}
\begin{barticle}
\bauthor{\bsnm{Dorey}, \binits{P.}},
\bauthor{\bsnm{Mersh}, \binits{K.}},
\bauthor{\bsnm{Romanczukiewicz}, \binits{T.}},
\bauthor{\bsnm{Shnir}, \binits{Y.}}:
\batitle{Kink-antikink collisions in the $\phi$ 6 model}.
\bjtitle{Physical review letters}
\bvolume{107}(\bissue{9}),
\bfpage{091602}
(\byear{2011})
\end{barticle}
\endbibitem

\bibitem[\protect\citeauthoryear{Gani et~al.}{2014}]{gani2014kink}
\begin{barticle}
\bauthor{\bsnm{Gani}, \binits{V.A.}},
\bauthor{\bsnm{Kudryavtsev}, \binits{A.E.}},
\bauthor{\bsnm{Lizunova}, \binits{M.A.}}:
\batitle{Kink interactions in the (1+ 1)-dimensional $\varphi$ 6 model}.
\bjtitle{Physical Review D}
\bvolume{89}(\bissue{12}),
\bfpage{125009}
(\byear{2014})
\end{barticle}
\endbibitem

\bibitem[\protect\citeauthoryear{Weigel}{2014}]{weigel2014kink}
\begin{bchapter}
\bauthor{\bsnm{Weigel}, \binits{H.}}:
\bctitle{Kink--antikink scattering in $\varphi$4 and $\phi$6 models}.
In: \bbtitle{Journal of Physics: Conference Series},
vol. \bseriesno{482},
p. \bfpage{012045}
(\byear{2014}).
\bcomment{IOP Publishing}
\end{bchapter}
\endbibitem

\bibitem[\protect\citeauthoryear{Bazeia et~al.}{2019}]{bazeia2019kink}
\begin{barticle}
\bauthor{\bsnm{Bazeia}, \binits{D.}},
\bauthor{\bsnm{Gomes}, \binits{A.R.}},
\bauthor{\bsnm{Nobrega}, \binits{K.}},
\bauthor{\bsnm{Simas}, \binits{F.C.}}:
\batitle{Kink scattering in a hybrid model}.
\bjtitle{Physics Letters B}
\bvolume{793},
\bfpage{26}--\blpage{32}
(\byear{2019})
\end{barticle}
\endbibitem

\bibitem[\protect\citeauthoryear{Belendryasova and Gani}{2019}]{belendryasova2019scattering}
\begin{barticle}
\bauthor{\bsnm{Belendryasova}, \binits{E.}},
\bauthor{\bsnm{Gani}, \binits{V.A.}}:
\batitle{Scattering of the $\varphi$8 kinks with power-law asymptotics}.
\bjtitle{Communications in Nonlinear Science and Numerical Simulation}
\bvolume{67},
\bfpage{414}--\blpage{426}
(\byear{2019})
\end{barticle}
\endbibitem

\bibitem[\protect\citeauthoryear{Gani et~al.}{2015}]{gani2015kink}
\begin{barticle}
\bauthor{\bsnm{Gani}, \binits{V.A.}},
\bauthor{\bsnm{Lensky}, \binits{V.}},
\bauthor{\bsnm{Lizunova}, \binits{M.A.}}:
\batitle{Kink excitation spectra in the (1+ 1)-dimensional $\varphi$8 model}.
\bjtitle{Journal of High Energy Physics}
\bvolume{2015}(\bissue{8}),
\bfpage{1}--\blpage{21}
(\byear{2015})
\end{barticle}
\endbibitem

\bibitem[\protect\citeauthoryear{Christ and Lee}{1975}]{christ1975quantum}
\begin{barticle}
\bauthor{\bsnm{Christ}, \binits{N.}},
\bauthor{\bsnm{Lee}, \binits{T.}}:
\batitle{Quantum expansion of soliton solutions}.
\bjtitle{Physical Review D}
\bvolume{12}(\bissue{6}),
\bfpage{1606}
(\byear{1975})
\end{barticle}
\endbibitem

\bibitem[\protect\citeauthoryear{Dorey et~al.}{2023}]{dorey2023collisions}
\begin{barticle}
\bauthor{\bsnm{Dorey}, \binits{P.}},
\bauthor{\bsnm{Gorina}, \binits{A.}},
\bauthor{\bsnm{Roma{\'n}czukiewicz}, \binits{T.}},
\bauthor{\bsnm{Shnir}, \binits{Y.}}:
\batitle{Collisions of weakly-bound kinks in the christ-lee model}.
\bjtitle{Journal of High Energy Physics}
\bvolume{2023}(\bissue{9}),
\bfpage{1}--\blpage{25}
(\byear{2023})
\end{barticle}
\endbibitem

\bibitem[\protect\citeauthoryear{Levkov et~al.}{2020}]{levkov2020chaotic}
\begin{barticle}
\bauthor{\bsnm{Levkov}, \binits{D.}},
\bauthor{\bsnm{Maslov}, \binits{V.}},
\bauthor{\bsnm{Nugaev}, \binits{E.Y.}}:
\batitle{Chaotic solitons in driven sine-gordon model}.
\bjtitle{Chaos, Solitons \& Fractals}
\bvolume{139},
\bfpage{110079}
(\byear{2020})
\end{barticle}
\endbibitem

\bibitem[\protect\citeauthoryear{Anninos et~al.}{1991}]{anninos1991fractal}
\begin{barticle}
\bauthor{\bsnm{Anninos}, \binits{P.}},
\bauthor{\bsnm{Oliveira}, \binits{S.}},
\bauthor{\bsnm{Matzner}, \binits{R.A.}}:
\batitle{Fractal structure in the scalar $\lambda$ ($\varphi$ 2- 1) 2 theory}.
\bjtitle{Physical Review D}
\bvolume{44}(\bissue{4}),
\bfpage{1147}
(\byear{1991})
\end{barticle}
\endbibitem

\bibitem[\protect\citeauthoryear{Kudryavtsev}{1975}]{kudryavtsev1975solitonlike}
\begin{botherref}
\oauthor{\bsnm{Kudryavtsev}, \binits{A.}}:
Solitonlike solutions for a higgs scalar field.
Technical report,
Institute of Theoretical and Experimental Physics
(1975)
\end{botherref}
\endbibitem

\bibitem[\protect\citeauthoryear{Adam et~al.}{2023}]{adam2023relativistic}
\begin{barticle}
\bauthor{\bsnm{Adam}, \binits{C.}},
\bauthor{\bsnm{Ciurla}, \binits{D.}},
\bauthor{\bsnm{Oles}, \binits{K.}},
\bauthor{\bsnm{Romanczukiewicz}, \binits{T.}},
\bauthor{\bsnm{Wereszczynski}, \binits{A.}}:
\batitle{Relativistic moduli space and critical velocity in kink collisions}.
\bjtitle{Physical Review E}
\bvolume{108}(\bissue{2}),
\bfpage{024221}
(\byear{2023})
\end{barticle}
\endbibitem

\bibitem[\protect\citeauthoryear{Fern{\'a}ndez-Mart{\'\i}nez and S{\'a}nchez-Granero}{2012}]{fernandez2012fractal}
\begin{barticle}
\bauthor{\bsnm{Fern{\'a}ndez-Mart{\'\i}nez}, \binits{M.}},
\bauthor{\bsnm{S{\'a}nchez-Granero}, \binits{M.}}:
\batitle{Fractal dimension for fractal structures: A hausdorff approach}.
\bjtitle{Topology and its Applications}
\bvolume{159}(\bissue{7}),
\bfpage{1825}--\blpage{1837}
(\byear{2012})
\end{barticle}
\endbibitem

\bibitem[\protect\citeauthoryear{G{\'o}rski et~al.}{2011}]{gorski2011accuracy}
\begin{botherref}
\oauthor{\bsnm{G{\'o}rski}, \binits{A.Z.}},
\oauthor{\bsnm{Drozdz}, \binits{S.}},
\oauthor{\bsnm{Mokrzycka}, \binits{A.}},
\oauthor{\bsnm{Pawlik}, \binits{J.}}:
Accuracy analysis of the box-counting algorithm.
arXiv preprint arXiv:1111.5749
(2011)
\end{botherref}
\endbibitem

\end{thebibliography}

\end{document}